\documentclass[review]{elsarticle}

\usepackage{lineno,hyperref}
\usepackage{graphicx}
\usepackage{txfonts}
\usepackage{amsfonts,amssymb,gensymb,algorithm,algpseudocode,lipsum}
\usepackage{natbib}

\journal{Advances in Space Research}

\biboptions{authoryear}

\begin{document}

\begin{frontmatter}

\title{Directed Energy Interception of Satellites}

\author[a,b]{Harrison She\corref{mycorrespondingauthor}}
\cortext[mycorrespondingauthor]{Corresponding author}
\ead{hshe55@gmail.com}

\author[b]{Will Hettel}
\author[b]{Phillip Lubin}
\address[a]{Electrical and Computer Engineering and Physics Departments, University of Auckland, Auckland 1010, New Zealand}
\address[b]{Physics Department, University of California, Santa Barbara, CA USA 93106-9530}

\begin{abstract}
High power Earth and orbital-based directed energy (DE) systems pose a potential hazard to Earth orbiting spacecraft. The use of very high power, large aperture DE systems to propel spacecraft is being pursued as the only known, feasible method to achieve relativistic flight in our NASA Starlight and Breakthrough Starshot programs. In addition, other beamed power mission scenarios, such as orbital debris removal and our NASA program using DE for powering high performance ion engine missions, pose similar concerns. It is critical to quantify the probability and rates of interception of the DE beam with the approximately 2000 active Earth orbiting spacecraft. We have modeled the interception of the beam with satellites by using their orbital parameters and computing the likelihood of interception for many of the scenarios of the proposed systems we are working on. We are able to simulate both the absolute interception as well as the distance and angle from the beam to the spacecraft, and have modeled a number of scenarios to obtain general probabilities. We have established that the probability of beam interception of any active satellite, including its orbital position uncertainty, during any of the proposed mission scenarios is low ($\approx10^{-4}$). The outcome of this work gives us the ability to predict when to energize the beam without intercept, as well as the capability to turn off the DE as needed for extended mission scenarios. As additional satellites are launched, our work can be readily extended to accommodate them. Our work can also be used to predict interception of astronomical adaptive optics guide-star lasers as well as more general laser use.
\end{abstract}


\end{frontmatter}

\section{Introduction}

\label{Introduction}

\subsection{Background}
\label{Background}

Directed energy systems are used routinely in astronomical application for laser guide-star excitation and for spacecraft laser communication. These systems typically have powers of 1--100 watts with sub-meter apertures. Larger scale directed energy systems have been proposed as a method of achieving relativistic flight \citep{pml2013,pml2014,pml2015,pml2016} to allow the first interstellar missions and rapid interplanetary transit. These DE systems are being actively pursued as a part of our NASA Starlight program, where a large aperture (1--10 kilometer diameter) phased array with multi GW power levels \citep{pml2016}. The derivative Breakthrough Starshot program uses the same technology, but focuses on the wafer-scale spacecraft and the ground-based array option. Laser systems have also been considered to characterize and intentionally remove orbital space debris in recent years \citep{Lejba20182609, Wang20161854, Hou20161698}. In all of these cases, the DE beam has sufficient flux to cause damage to active Earth-orbiting satellites, which may be inadvertently illuminated. \\

To be able to accurately simulate such scenarios, it is imperative that the orbits of the satellites are determined accurately in order to predict when they may potentially intercept the beam.  Space Situational Awareness (SSA) is concerned with the monitoring of the many Earth-orbiting bodies. Such programs are run both by the United States (NASA and DoD) and the European Space Agency (ESA). These programs involve the observation of the space environment, and the identification and tracking of space objects in that environment for international safety and security \citep{rovetto2016preliminaries}. Crucially, SSA data from the North American Aerospace Defense Command (NORAD) can be acquired through Space Track, or an orbital body bulletin system called CelesTrak \citep{kelso}. This data aids in the tracking and ephemeris generation of satellite bodies in orbit around the Earth by providing two-line element sets (TLE) which encode a list of orbital elements to predict the position of each Earth-orbiting object at a specific time. In this paper we detail the development and use of simulation code to estimate the probability of directed energy beam interception with an active satellite. Simulations are performed for a number of scenarios that pertain to future planned missions to Alpha Centauri, the Moon, Mars and Pluto. We first motivate the endeavour, and follow this with an analytical approximation to derive expected probabilities with which to compare our simulation results. This is followed by a detailed explanation of the simulation and the resulting outcomes and conclusions that may be drawn.

\subsection{Motivation}
\label{Motivation}

We use TLEs to simulate, analyze, and evaluate the frequency and duration of potential intercepts with the DE beam, as well as the distance by which an active satellite will miss the beam, when energized from a given location, at a given direction (both celestial and terrestrial) and time. Previously, the prediction of the orbits of satellites for beam intercept purposes was left to agencies that would give a ``go/no-go'' response, which is not adequate for understanding the operational implications of a system such as ours, nor does it allow for the precise orbital information necessary for an understanding of required optical sidelobe suppression. We use the term ``intercept'' to generally refer to the beam crossing the estimated position of the satellite including its positional uncertainty. Gating the DE off during all intercepts will ensure that no satellites are illuminated. The positional uncertainty of satellites is typically on the scale of kilometers \citep{revisiting} and is thus vastly larger than the physical satellite. The actual probability of direct illumination of the physical satellite with the beam is significantly lower than with the positional uncertainty of the satellite; we return to this later. Our interception calculations allow us to answer a number of questions including:

\begin{itemize}

\item How likely is a DE beam to intercept an active satellite, including its orbital uncertainty, during any given mission?
\item How many satellites are likely to be intercepted by the DE beam for a given mission scenario?
\item How many times, and for what duration, will the DE have to be gated off, and hence what is the transmission proportion that is likely to be achieved for a given DE mission?
\end{itemize} 

Simulation of such outputs allows us to gauge the feasibility of a number of future directed energy missions, such as those proposed by \citet{pml2016}. These missions can vary in total DE beaming duration from hundreds of seconds to several years. Shorter DE exposure missions include the proposal to accelerate ultra-light wafer spacecraft to relativistic velocities to reach Alpha Centauri. Longer DE exposure missions include, but are not limited to, continuous DE propulsion of larger scale payloads to closer range targets such as Mars, as well as missions to the moon or other planets. The results obtained from this research can also be applied to adaptive optics applications and other beamed power applications. \\

In our calculations and simulations, we choose to only account for intercepts with active satellites, as we are not concerned with damaging inactive satellites or space debris. These objects will not significantly impact our missions by crossing the beam because they are typically small compared to the beam, and the intercept durations will be short (discussed in Sec. \ref{Analytical Approximation}). Reflections off of these objects will be largely isotopic and thus low intensity at distance. See Sec. \ref{Directed energy safety concerns} for expected beam intensity. \\

This paper discusses the design and development of a DE array and satellite ephemeris interdiction simulator. This simulator reads TLE objects from an online satellite catalog and subsequently calculates the ephemerides of active satellites that may be damaged during DE transmission. With the aid of the \texttt{PyEphem} library, which provides useful generic astronomical calculation functions, the aforementioned ephemerides can subsequently be utilized to simulate orbital motion \citep{rhodes}. These, in conjunction with laser array model parameters, can be used to calculate potential intercepts with the DE beam.  The simulator program allows the user to input a specific laser array site, a pointing target (right ascension, declination), a beam diameter, and start and end times for each simulation. With these parameters, the program calculates the total interception time, the total time the DE beam can transmit, and the distance (in linear and angular terms) between the DE beam and each satellite on the array's horizon. The program also includes a job scheduler that can be used to run multiple simulations in order to rapidly evaluate a multitude of potential mission scenarios that may arise. 

\section{Analytical approximation}
\label{Analytical Approximation}

In this section, we analytically calculate the results we expect to obtain from our simulations. We use a simplified model of the Earth and its satellites to produce rough, order of magnitude estimates of intercept probability, beam crossing duration, and intercept frequency. The analytical results are intended to provide a intuitive baseline from which to compare the simulation results.\\

We can approximate the laser-satellite intercept probability by assuming that the satellite distribution is isotropic at each altitude. This assumption allows us to estimate the intercept probability distribution at any location on Earth. We assume that satellites are ``points'' with an error ``sphere'' due to position determination uncertainty and allow for either a diverging or converging beam, with laser array (and hence beam) diameter $d$. For simplicity we assume the Earth is perfectly spherical and the satellite orbits are circular.\\

It is natural to expect that the number of interceptions with the DE beam should be low, since the distribution of Earth-orbiting spacecraft is sparse. If all 16776 satellite elements from the Space Track database were on the Earth's surface, there would be less than one per 30~thousand~$\mathrm{km}^2$. \\

\begin{table*}[ht!]
\centering
\scriptsize
\caption{Definition of Constants and Variables Used in Section 2}\label{tab:pluto}
 \begin{tabular}{c c c}
 \hline\hline
Symbol & Definition & Value/Units  \\ 
 \hline
 $\delta_{\mathrm{sat}}$ & diameter of satellite position uncertainty & km  \\ 

 $R_{{\oplus}}$ & radius of Earth & 6371 km \\ 

 $h$ & height of a satellite above sea level & km \\ 

 $d$ & diameter of laser optics & m or km \\ 
 
 $\theta_{1/2}$ & laser divergence or convergence half-angle & radians \\
 
 $A_\mathrm{intercept}$ & effective area for a DE beam to intercept a satellite & $\mathrm{km}^2$ \\
 
 $d_\mathrm{sat}$ & actual size of satellite $<< \delta _{\mathrm{sat}}$ & m \\
 
 $n_{\mathrm{int-inst}}(h)$ & instantaneous number of beam intercepts with a single satellite at height $h$ & dimensionless \\
 
 $n$ & total number of satellites & dimensionless \\
 
 $N_\mathrm{int-inst}$ & instantaneous number of beam intercepts with all satellites & dimensionless \\
 
 $\lambda$ & wavelength of laser light & nm \\
 
 $F_\mathrm{c}$ & centripetal force acting on a satellite in orbit & N \\
 
 $F_\mathrm{g}$ & gravitational force acting on a satellite in orbit & N \\
 
 $M_\mathrm{sat}$ & mass of satellite & kg \\
 
 $v(h)$ & speed of satellite at height $h$ & m $\mathrm{s}^{-1}$ \\
 
 $M_\oplus$ & mass of Earth & $5.972 \times 10^{24}$ kg \\
 
 $G$ & gravitational constant & $6.67408 \times 10^{-11} \ \mathrm{m}^3 \ \mathrm{kg}^{-1} \ \mathrm{s}^{-2}$ \\
 
 $t_{\mathrm{int-inst}}(h)$ & worst-case duration of single satellite beam crossing including $\delta_\mathrm{sat}$ & s \\
 
 $\frac{\delta n_{\mathrm{\mathrm{int}}} }{\delta t}(h)$ & number of single satellite intercepts per unit time & $\mathrm{s}^{-1}$ \\
 
 $\frac{\delta N_{\mathrm{\mathrm{int}}} }{\delta t}$ & total number of satellite intercepts per unit time & $\mathrm{s}^{-1}$ \\
 
 $t_{\mathrm{int-sat}}(h)$ & worst-case duration of single satellite beam crossing not including $\delta_\mathrm{sat}$ & s \\
 
 \hline
\end{tabular}
\end{table*}

\subsection{Intercept Probability}
\label{Intercept probability}

At the time of writing this paper, the full catalog of TLEs retrieved from the Space Track database consisted of 16776 satellite elements, most larger than 10~cm diameter. These elements comprise unclassified active and inactive satellites, as well as satellite debris. For our purposes, we are only concerned with the 1783 active satellite elements; this also includes GEO spacecraft for added simulation robustness. We also do not worry about inactive satellites or debris since beam interception of these objects do not pose a threat. It is important to also note that at least 100 new satellites are launched into space each year (which will increase in the future), and hence it is important to update the TLE catalog used on a regular basis to include any new satellites \citep{finkleman2014dilemma}. \\ 
                                                                                                                                                                                                                                                                                                                                                                                                                                                                                                                                                                                                                                                                                                                                                                                                                                                                                                                                                                                                                                                                                                                                                                                                                                                                                                                                                                                                                                                                                                                                                                                                                                                                                                                                                                                                                                                                                                                                                                                                                                                                                                                                                                                                                                                                                                                                                                                                                                                                                                                                                                                                                                                                                                                                                                                                                                                                                                                                                                                                                                                                                                                                                                                                                                                                                                                                                                                                                                                                                                                                                                                                                                                                                                                                                                             
\begin{figure}[ht!]
\centering
\includegraphics[width=1\linewidth]{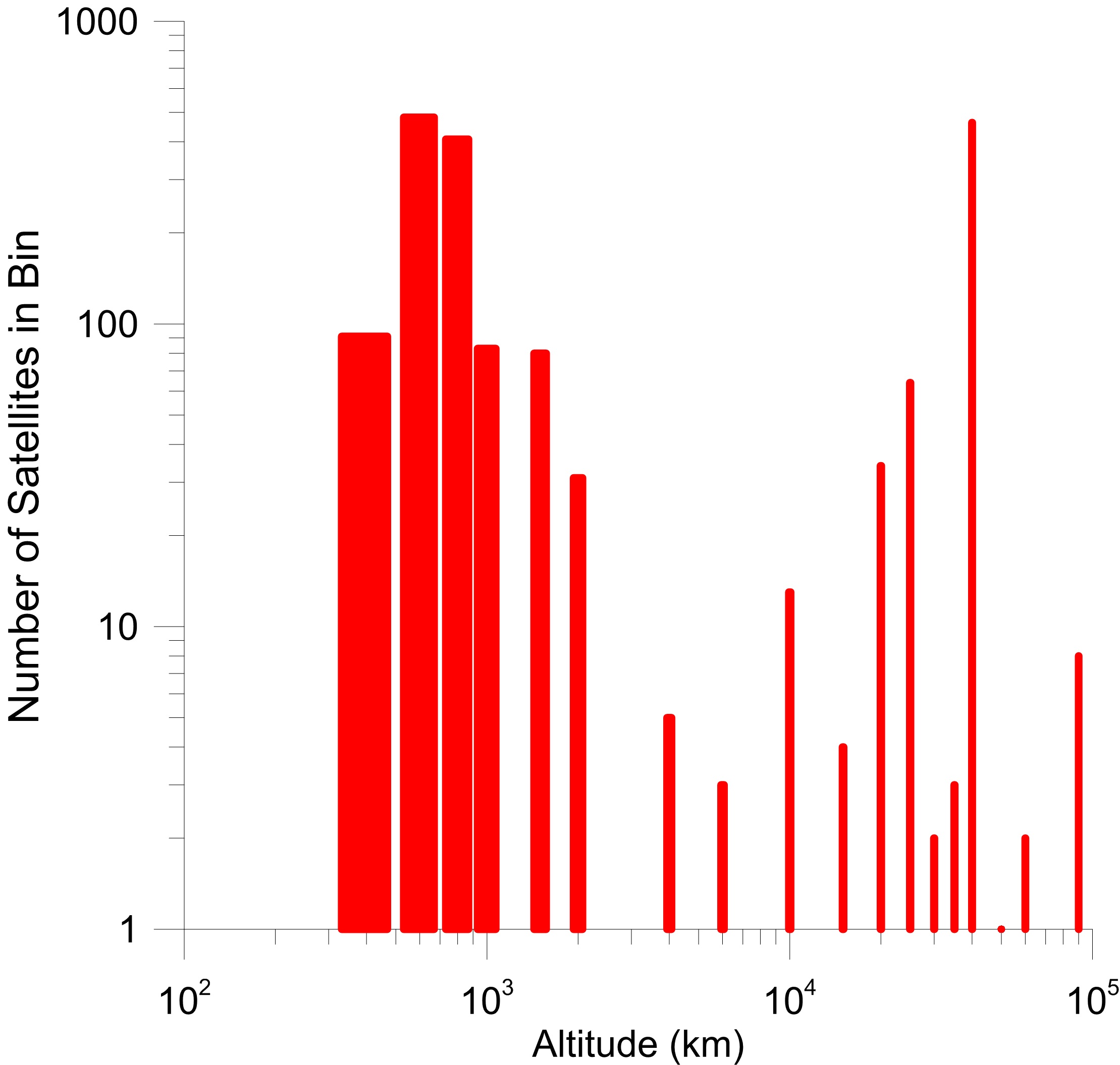} \caption{The distribution of 1783 active satellite altitudes at a given time. Most satellites fall into either Lower Earth Orbits (LEO) or Geosynchronous Orbits (GEO). Due to the dynamic range and nonuniformity in the distribution of satellite altitudes, the size of each bin increases with altitude, starting at 100~km and ending at 50~thousand~km.} 
\label{fig:hist}
\end{figure}

Using \texttt{PyEphem}, an astronomical python package that provides powerful generic astronomical functions, the satellite altitudes from sea level have been calculated at an arbitrary point in time (midnight 2018/2/15) and plotted as a distribution seen in Fig. \ref{fig:hist}. From this plot, if we assume that most orbits are roughly circular (the median eccentricity is $6\times 10^{-3}$ for all 16776 TLE elements and $7\times 10^{-4}$ for the 1783 active satellites, as calculated using the same TLE data), we can see that most satellites fall into either the Low Earth or Geostationary orbital regimes. \\

In order to predict the probability of intercepting any satellite at a given time, we can use a simplified physical model detailed below to estimate approximate figures for our simulation. The simulation can also provide a means of validation for our model.\\

Let us assume that the Earth is a perfect sphere with a radius of 6371 km and we aim the DE beam in the zenith direction (directly upwards). If using the SGP4 model, which is used by \texttt{PyEphem}, the satellites deviate from their idealized orbits described by their TLE files by 1--3~km per day \citep{satuncertainty}. We can use this information to model the tolerance (intercept volume) of each satellite as a 6~km diameter sphere, therefore setting $\delta_{\mathrm{sat}} = 6$~km. We assume that the TLEs can be updated on a daily basis so $\delta_\mathrm{sat}$ does not increase with time. The DE beam can be modeled as a cone extending from Earth to a geocentric sphere with radius $R_{\oplus}+h$. Setting the initial diameter of this cone to $d$ and the slope of its sides to the laser convergence/divergence half-angle $\theta_{1/2}$, the diameter of the DE beam at a given height is $d + 2h\theta_{1/2}$. $\theta_{1/2}$ is positive for a diverging beam and negative for a converging beam.\\

\begin{figure}[ht!]
\centering
\includegraphics[width=1\linewidth]{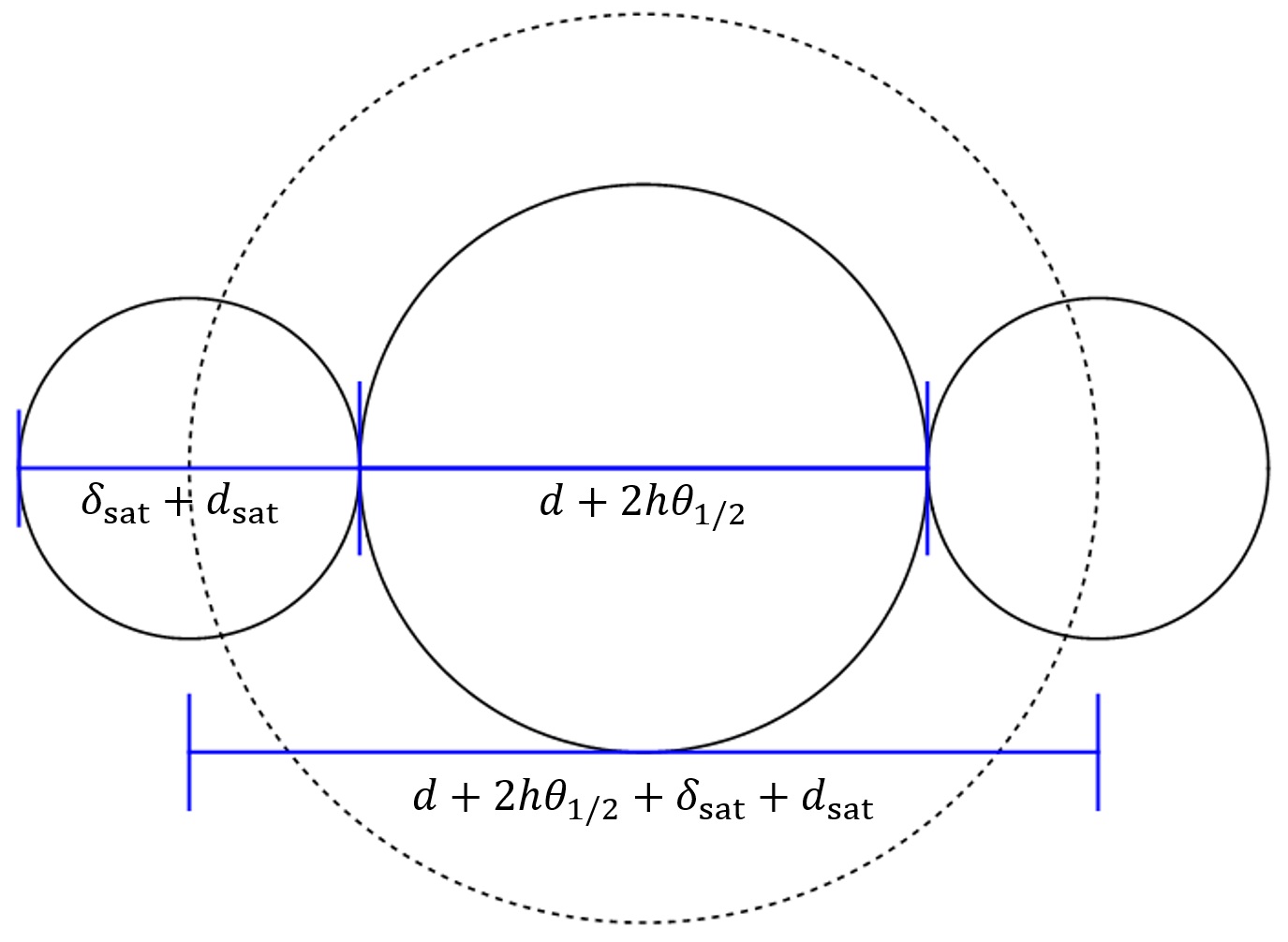} \caption{A model showing a DE beam (center) and two edge cases for intercepts (sides). An intercept is counted if the coordinate of satellite according to a TLE file (located at the center of the satellite's error sphere) is within $(\delta_\mathrm{sat}+d_\mathrm{sat})/2$ of the edge of the beam, which ensures that all cases that a satellite may be illuminated are accounted for. This condition creates an effective ``intercept area'' equal to that of a circle of diameter $d + 2h\theta_{1/2} + \delta_\mathrm{sat} + d_\mathrm{sat}$. Intercepts are counted when TLE files report coordinates that are within this area.} 
\label{fig:hitbox}
\end{figure}

We model the distribution of the satellites as uniform at any given moment and calculate the probability of any single given satellite hitting the beam as the ratio of ``intercept area'' (shown in Fig. \ref{fig:hitbox}) to the total area of the geocentric sphere:

\begin{eqnarray}
n_{\mathrm{int-inst}}(h) &=& \frac{A_{\mathrm{Intercept}}}{4\pi (R_{\oplus}+h)^2}\\
&=& \frac{\pi [d+2h\theta _{1/2} + \delta _{\mathrm{sat}} +d_{\mathrm{sat}} ]^{2} /4}{4\pi (R_{\oplus} +h)^{2} }\\
&=& \frac{[d+2h\theta _{1/2} + \delta _{\mathrm{sat}} +d_{\mathrm{sat}} ]^{2} }{16(R_{\oplus} +h)^{2} }.
\end{eqnarray}
A plot of $n_{\mathrm{int-inst}}$ as a function of $h$ is shown in Fig. \ref{fig:las}. By summing $n_{\mathrm{int-inst}}$ for $n$ satellites, we find the probability that the beam will intercept any satellite at the given time is

\begin{equation} 
%\begin{split}
N_\mathrm{int-inst} = \sum_{i=1}^{n}\frac{[d+2h_i\theta _{1/2} + \delta _{\mathrm{sat}} +d_{\mathrm{sat}} ]^{2} }{16(R_{\oplus} +h_i)^{2} }.
%\end{split}
\end{equation}

We now assume that the diameter of the satellite and the laser half-angle are negligible ($d_{\mathrm{sat}} = \theta_{1/2} = 0$) and calculate $N_\mathrm{int-inst}$ for a laser array with $d = 10~\mathrm{km}$ and $\delta_{\mathrm{sat}} = 6~\mathrm{km}$ and for $n = 1783$ to only account for active satellites. We are justified in setting $\theta_{1/2}=0$ because proposed missions require focusing at high altitude, so a significant divergence or convergence will not be expected. Furthermore, the half-angle due to diffraction ($\sim\lambda/d$) of the 10~km laser array of 1064~nm wavelength proposed by \citet{pml2016} is on the order of $10^{-10}$~radians and is thus negligible compared to $d/h$ and $\delta_\mathrm{sat}/h$.

We then have

\begin{equation}
%\begin{split}
N_\mathrm{int-inst} = 3.9 \times 10^{-4}.
%\end{split}
\end{equation}

Therefore, from this result, we can see that even if no specific precautions are taken, the likelihood of an intercept in this conservative configuration is less than $10^{-3}$. Note that the likelihood of a satellite actually being illuminated by the beam is smaller by a factor of $(\frac{d+d_\mathrm{sat}}{d+\delta_\mathrm{sat}})^2$ ($\approx 0.4$ for $d=10$~km and $\delta_\mathrm{sat} = 6$~km), because the true diameter of the satellite is $d_\mathrm{sat}$ as opposed to $\delta_\mathrm{sat}$. One should also consider that the effective diameter of the beam at altitude will often be less than $d$ (depending on focusing requirements for missions), further reducing the probability of illumination. For demonstrative purposes, if we pick a diameter of 3~m for all satellites, we can find the actual area covering fraction of satellites in the sky by setting $d = \delta_\mathrm{sat} = 0$:

\begin{equation}
%\begin{split}
N_\mathrm{int-inst} = 1.4 \times 10^{-11}\\
%\end{split}
\end{equation}
for $d_\mathrm{sat} = 3$~m and $n = 1783$. There is nothing special about our specific choice of satellite diameter other than that it yields a rough estimate. As we can see, the true covering fraction of active satellites is smaller than that which includes $d$ and $\delta_\mathrm{sat}$ by many orders of magnitude.

\subsection{Satellite Beam Crossing Duration and Frequency}
\label{Satellite beam crossing duration and frequency}

It is important to not only characterize the transmission ratio, but also to characterize the duration and frequency of events in which the DE must be gated off. We can calculate these quantities using the velocities of spacecraft. If we again assume that orbits are circular and set a satellite's centripetal force $F_c$ equal to its gravitational force $F_g$, we have

\begin{eqnarray}
F_\mathrm{c} &=& \frac{M_{\mathrm{sat}} \ v(h)^2}{R_{\oplus} + h}\\
F_\mathrm{g} &=& \frac{G M_{\mathrm{sat}} M_{\oplus}}{(R_{\oplus} + h)^2}\\
\mathrm{setting} \ F_\mathrm{c}  = F_\mathrm{g} \\
v(h) &=& \sqrt{\frac{G M_{\oplus}}{R_{\oplus} + h}}.
\end{eqnarray}
A plot of $v$ as a function of $h$ is shown in Fig. \ref{fig:las}. \\

We obtain the mean velocity of all satellites by summing their individual velocities and dividing by the number of satellites:

\begin{eqnarray}
\overline{v} &=& \frac{1}{n}\sum_{i=1}^{n}\sqrt{\frac{G M_{\oplus}}{R_{\oplus} + h_i}}\\
&=& 6086 \ \mathrm{m \ s}^{-1}
\end{eqnarray}
for the 1783 satellites used from the TLE data available to us. The worst-case duration of an intercept will be when the expected position of the satellite passes through the center of the beam. Since it takes a full day for a satellite to deviate from its expected orbit by 3~km, we assume that the velocity described by the TLE file is approximately equal to that of the satellite. We can thus assume that the distance traveled by the satellite during $t_{\mathrm{int-inst}}$ is the same as that of its error sphere. We calculate $t_{\mathrm{int-inst}}$ for a stationary DE beam with $d = 10$~km and $\theta_{1/2}=0$, and a satellite with $\delta_\mathrm{sat} = 6$~km and $d_\mathrm{sat}=0$ moving at average velocity to be

\begin{eqnarray}
t_{\mathrm{int-inst}}(h,\overline{v}) &=& \frac{d+2h\theta _{1/2} +\delta _{\mathrm{sat}} +d_{\mathrm{sat}}}{\overline{v}}\\ 
&=& 2.63 \ \mathrm{s}.
\end{eqnarray}
Figure \ref{fig:las} shows $t_\mathrm{int-inst}$ as a function of $h$. We will now calculate the number of intercepts with a given satellite per unit time. This will be equal to the area swept out by the satellite-laser system per unit time, divided by the area of the geocentric sphere of radius $R_{\oplus}+h$:

\begin{eqnarray}
\frac{\delta n_{\mathrm{int}} }{\delta t}(h) 
&=& \frac{(d+2h\theta _{1/2} +\delta _{\mathrm{sat}} +d_{\mathrm{sat}}){v(h)}}
{4 \pi{(R_{\oplus} +h)}^2 } \\
&=& \frac{{4v}n_{\mathrm{int-inst}}(h) }{\pi (d+2h\theta _{1./2} +\delta _{\mathrm{sat}} +d_{\mathrm{sat}} )} \\
&=& \frac{4n_{\mathrm{int-inst}}(h) }{\pi {t}_{\mathrm{int-inst}}(h) }.
\end{eqnarray}
A plot of $\frac{\delta n_{\mathrm{int}} }{\delta t}$ as a function of $h$ is shown in Fig. \ref{fig:las}. For all satellites, we can sum the number of intercepts per unit time to acquire a general intercept frequency:

\begin{eqnarray}
%\begin{split}
\frac{\delta N_{\mathrm{int}}}{\delta t} &=& \sum \frac{\delta n_{\mathrm{int}} }{\delta t}(h) = \frac{4}{\pi} \sum_{i=1}^{n} \frac{n_{\mathrm{int-inst}}(h_i) }{{t}_{\mathrm{int-inst}}(h_i) }\\
\frac{\delta N_{\mathrm{int}}}{\delta t} &=& 0.000232 \ \mathrm{s}^{-1}
%\end{split}
\end{eqnarray}
for $d = 10$~km, $\theta_{1/2} = 0$, $\delta_\mathrm{sat} = 6$~km, $d_\mathrm{sat} = 0$ and $n = 1783$. Thus, from this analysis we can see that the estimated intercepts per second is very low, and only once every 4310 seconds is it expected that a satellite will intercept the beam.\\

We will now calculate the worst-case intercept duration per unit time. For a single satellite, this is simply its intercept frequency multiplied by $t_{\mathrm{int-inst}}$:

\begin{equation}
%\begin{split}
\frac{\delta n_{\mathrm{int}}}{\delta t}(h)t_{\mathrm{int-inst}}(h) = \frac{4}{\pi}n_{\mathrm{int-inst}}(h).
%\end{split}
\end{equation}

For all satellites, we have

\begin{eqnarray}
\sum \frac{\delta n_{\mathrm{int}}}{\delta t}(h)t_{\mathrm{int-inst}}(h) &=& \frac{4}{\pi}\sum_{i=1}^{n} n_{\mathrm{int-inst}}(h_i)\\
&=& \frac{4}{\pi}N_\mathrm{int-inst} \\
&=& 4.9 \times 10^{-4}
\label{wc int fraction}
\end{eqnarray}
for $d = 10$~km, $\theta_{1/2} = 0$, $\delta_\mathrm{sat} = 6$~km, $d_\mathrm{sat} = 0$, and $n = 1783$. \\

The values of intercept probability and frequency will certainly increase as the number of active satellites increases. This concern is particularly relevant to current plans to deploy large constellations of spacecraft. However, on the assumption that these values will scale linearly with the number of active satellites, the number of active satellites will need to increase by 2-3 orders of magnitude before significantly impacting the transmission time of DE missions.

\subsection{Actual Satellite Beam Crossing Time}

\begin{figure}[ht!]
\centering
\includegraphics[width=0.9\linewidth]{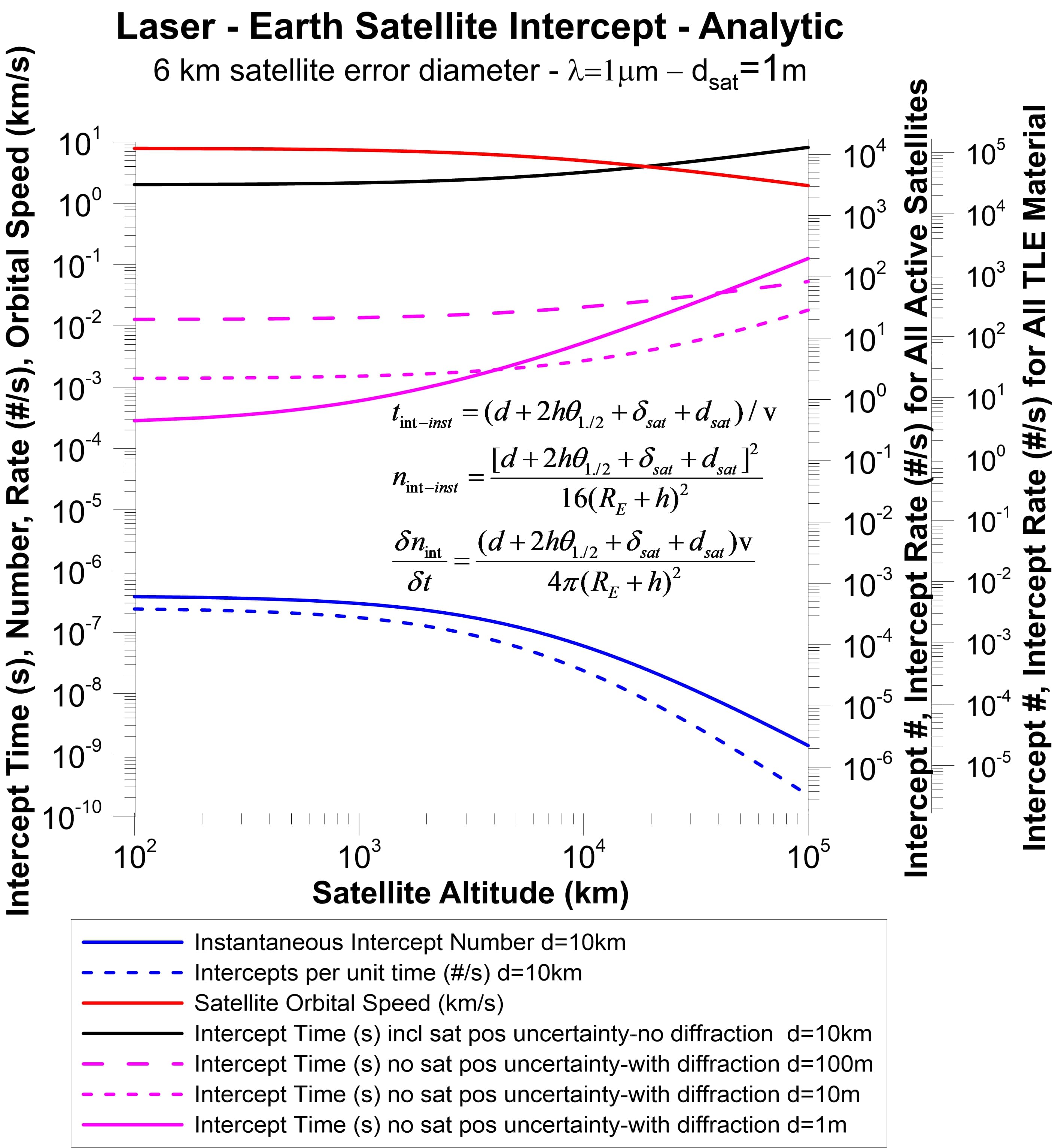} 
\caption{Intercept time, intercept rate, and orbital speed plotted as a function of satellite altitude. Here $\delta_\mathrm{sat} = 6$~km, $\lambda = 1064~\mathrm{nm}$, and $d_{\mathrm{sat}}=1~\mathrm{m}$. The solid blue line shows $n_\mathrm{int-inst}$, the red line shows $v$, the dotted blue line shows $\frac{\delta n_{\mathrm{int}} }{\delta t}$, the black line shows $t_{\mathrm{int-inst}}$, and the magenta lines show $t_{\mathrm{int-sat}}$ for various values of $d$. $n_\mathrm{int-inst}$ and $\frac{\delta n_{\mathrm{int}} }{\delta t}$ should be read on the right axes. $v$, $t_{\mathrm{int-inst}}$, and $t_{\mathrm{int-sat}}$ should be read on the left axis. The first right axis is scaled to account for all unclassified active satellites recognized by NORAD. The second right axis is scaled to account for all unclassified NORAD objects, including active and inactive satellites, debris, second stages, etc. From the various curves, one can see that the number of intercepts, the rate at which they occur, and their duration are expected to be modest.} 
\label{fig:las}
\end{figure}

The typical size of satellites is on the order of meters, as opposed to the uncertainty of the orbital position, which is on the order of kilometers. For some applications, such as astronomical observations, the total duration that the satellite is exposed to the beam may be important. The actual worst-case beam crossing duration is

\begin{equation}
t_{\mathrm{int-sat}}(h) =\frac{d+2h\theta _{1/2} +d_{\mathrm{sat}}}{v(h)}.
\end{equation}
Plots of $t_{\mathrm{int-sat}}$ as a function of $h$ for various values of $d$ are shown in Fig. \ref{fig:las}. In these plots, the laser divergence half-angle due to diffraction is included as it is not negligible for small aperture diameters. We assume a circular aperture, which has $\theta _{1/2} =1.22\lambda /d$.

\section{Directed energy safety concerns}
\label{Directed energy safety concerns}

In the event that the directed energy beam does in fact illuminate a satellite, it is important to note that for the mission to beam power to the moon, even for a 10~km laser array at the desired power of 100~GW, the flux is only approximately 1 $\mathrm{kW/m}^{2}$. This flux is about the same as sunlight at sea level. Satellites will thus not be destroyed by the beam, however optical sensors may be adversely affected if exposed for extended periods of time. A flux of approximately 100 $\mathrm{kW/m}^{2}$ warrants concern of thermal issues. For the interstellar mission to launch spacecraft to Alpha Centauri we can expect a much greater flux, as the beam needs to be focused on the wafer-scale spacecraft. Consequently, this is the most dangerous scenario for Earth-orbiting satellites. On the other hand, the interstellar case uses the DE beam for a much shorter period of time (at most 1000~s), so interceptions and instances where ground control needs to gate off the DE beam will be minimal. \\

Nonetheless, it is crucial that all directed energy missions obtain the authorization of the Laser Clearing House, as mandated by Department of Defense (DOD) Instruction 3100.11: Illumination of Objects in Space by Lasers \citep{mckeon_2016}.

\section{Methods} 
\label{Method}

\begin{table*}[ht!]
\centering
\scriptsize
\caption{Definition of Constants and Variables Used in Section 4}\label{tab:pluto}
 \begin{tabular}{c c c}
 \hline\hline
Symbol & Definition & Value/Units  \\ 
 \hline
 $p$ & distance from satellite to center of DE beam & km  \\ 

 $\phi$ & latitude of DE system & degrees \\ 

 $\lambda$ & longitude of DE system & degrees \\ 
 
 $\alpha$ & altitude angle of DE beam & degrees \\
 
 $\beta$ & azimuth angle of DE beam & degrees \\
 
 $\phi_\mathrm{s}$ & latitude of satellite & degrees \\
 
 $\lambda_{\mathrm{s}}$ & longitude of satellite & degrees \\
 
 $\alpha_\mathrm{s}$ & altitude angle of satellite from DE system site & degrees \\
 
 $\beta_\mathrm{s}$ & azimuth angle of satellite from DE system site & degrees \\
 
 $h$ & height of a satellite above sea level & km \\ 
 
 $r_k$ & distance from DE system site to satellite & km \\
 
 $R_{{\oplus}}$ & radius of Earth & 6371 km \\ 
 
 $\gamma$ & angle between DE system and satellite from center of Earth & degrees \\
 
 $\theta$ & angle between DE beam and satellite from DE system site & degrees \\
 
 $d$ & diameter of laser optics & m or km \\ 
 
 $\delta_{\mathrm{sat}}$ & diameter of satellite position uncertainty & km  \\ 
 
 $\Delta t$ & time between simulation points & s \\
 
 $v_\mathrm{LEO}$ & typical speed of a satellite in low Earth orbit & 7800 m $\mathrm{s}^{-1}$ \\
 
 $N_\mathrm{res}$ & number to determine the time resolution of the simulation & dimensionless \\
 
 $\delta t_\mathrm{int}$ & simulation error of duration of a single intercept & s \\
 
 $T_\mathrm{int}$ & duration of all intercepts & s \\
 
 $N_\mathrm{int}$ & number of intercepts with all satellites & dimensionless \\
 
 $\delta T_\mathrm{int}$ & simulation error of duration of all intercepts & s \\
 
 $\delta N_\mathrm{int}$ & simulation error of number of intercepts with all satellites & dimensionless \\
 
 $p_\mathrm{min}$ & satellite's closest distance of approach to center of DE beam & km \\
 
 $v$ & speed of satellite & m $\mathrm{s}^{-1}$ \\
 
 $A_\mathrm{intercept}$ & effective area for a DE beam to intercept a satellite & $\mathrm{km}^2$ \\
 
 $n_\mathrm{int}$ & number of intercepts with a single satellite & dimensionless \\
 
 $\delta n_\mathrm{int}$ & simulation error of number of intercepts with a single satellite & dimensionless \\
 
 c & speed of light & $2.998 \times 10^8 \ \mathrm{m \ s}^{-1}$ \\
 
 \hline
\end{tabular}
\end{table*}

\subsection{Dataset}

In an attempt to answer our proposed research questions, a simulator program was written in Python using the \texttt{PyEphem} library \citep{rhodes}. In order to model all the ephemerides of Earth-orbiting satellites, this library requires the parsing of TLEs for every unclassified tracked satellite we know about. \\ 

TLEs are fundamentally general perturbation element sets that contain mean values of orbital parameters, which are pre-processed by NORAD (North American Aerospace Defense Command) to remove periodic variations. Therefore, in order to use these TLEs to obtain accurate predictions, we must reconstruct the periodic variations in the appropriate manner that NORAD first removed these variations \citep{spacetrackReport, osweiler2006covariance}. The \texttt{PyEphem} astronomical python library gives us the ability to do this, as it utilizes SGP4 (Standard General Perturbations Satellite Orbit Model 4) to compute the ephemeris of each satellite defined by a TLE. \\

Such data can be retrieved from the Space Track or the CelesTrak online bulletin board system, which both acquire these TLEs directly from NORAD. CelesTrak also groups sets of TLEs into logical categories such as low Earth orbit (LEO), geo-synchronous orbit, scientific, communication, navigation, and more. \\
 
It is important to note that the SGP4 model has limited accuracy when used to derive the ephemeris of satellites. There exists inherent limitations in using TLE files for any analysis, since the calculation accuracy is limited by the number of decimal places that orbital parameter fields can fit in the 69 column TLE. As a direct result, TLE data can only be accurate to approximately 1--3~km at the time of the epoch, and degrade as calculations are performed further from this date \citep{satuncertainty,revisiting,sgp4,finkleman2014dilemma}. This uncertainty has been taken into account in both our analytical approximation and numerically simulated results. \\

The physical and mathematical method of how the ephemeris is derived from this model is presented in detail in Sec. \ref{Physics and mathematics of calculations}. In addition, how the ephemerides are subsequently processed in our simulation code is explained in Sec. \ref{Simulations}.

\subsection{Physics and Mathematics of Calculations}
\label{Physics and mathematics of calculations}

The SGP4 orbit propagator uses an analytic low-order solution to Newton's second law, giving a realistic model for gravitational potential and a dissipative atmospheric environment \citep{atmos}. This model is used for near-Earth satellites and was developed in 1970 by Ken Cranford \citep{lane1979general}, and is a simplification of the more complex theory of Lane and Cranford, which uses a power density function for the atmosphere and a gravitational model from Brouwer's solution \citep{brouwer1959solution}. Outlined below is a summary of the results deriving the position of a satellite from columns 09-63 of line 2 of each TLE, which use the following orbital elements at the Epoch: the mean motion ($n_o$), eccentricity ($e_o$), inclination ($i_o$), argument of perigee ($\omega_o$), the longitude of ascending node ($\Omega_o$), the mean anomaly ($M_o$), the first time derivative of the mean motion ($\dot{n}_o$), and the second time derivative of mean motion ($\ddot{n}_o$), to calculate the position and velocity vectors from the observer to a given satellite in the radial direction as ${\textit{\textbf{r}}}$ and $\dot{{\textit{\textbf{r}}}}$, respectively, using the method used in Space Track \citep{spacetrackReport}:

\begin{equation}
{\textit{\textbf{r}}}=r_k{\textit{\textbf{U}}}\
\end{equation}

\begin{equation}
\dot{{\textit{\textbf{r}}}}=\dot{r}_k{\textit{\textbf{U}}}+(r\dot{f})_k{\textit{\textbf{V}}},
\end{equation}
where
\begin{equation}
{\textit{\textbf{U}}}={\textit{\textbf{M}}}\sin u_k+{\textit{\textbf{N}}}\cos u_k
\end{equation}
\begin{equation}
{\textit{\textbf{V}}}={\textit{\textbf{M}}}\cos u_k-{\textit{\textbf{N}}}\sin u_k
\end{equation}
and
\[{\textit{\textbf{M}}}=\left\{\begin{array}{l}
                          M_x=-\sin\Omega_k\cos i_k\\
                          M_y=\cos\Omega_k\cos i_k\\
                          M_z=\sin i_k
                         \end{array}\right\}\]
\[{\textit{\textbf{N}}}=\left\{\begin{array}{l}
                          N_x=\cos\Omega_k\\
                          N_y=\sin\Omega_k\\
                          N_z=0
                         \end{array}\right\}\]

\[r_k=r\left[1-\frac32k_2\frac{\sqrt{1-e_\mathrm{L}{}^2}}{p_\mathrm{L}{}^2}(3\theta^2-1)\right]
+\Delta r\]
\[u_k=u+\Delta u\]
\[\Omega_k=\Omega+\Delta\Omega\]
\[i_k=i_o+\Delta i.\]
Symbols not defined here can be viewed alongside the complete derivation of these results from the Space-Track report in \ref{Appendix} \citep{spacetrackReport}. 
                     
\subsection{Simulations}
\label{Simulations}

\subsubsection{Calculating total intercept duration}
\label{Calculating total intercept duration}

A single TLE can be read in through \texttt{PyEphem}'s \texttt{readtle()} function, which creates a body object for that satellite segment. Subsequently, these satellite objects are used to compute ephemeris, model the laser and calculate the intercepts. At a high level, this can be described in the following steps:

\begin{enumerate}
\itemsep=0em
 \item Retrieve all satellites\textsc{\char13} orbital parameters as TLEs (two-line element sets) from the Space Track database (NORAD). 
  \item Use \texttt{ephem.compute} to derive the position (altitude angle, azimuth angle, latitude, longitude, and elevation) for each satellite at a particular date and time.
  \item Model the tolerance (intercept volume) of each satellite using the information that satellites deviate from their idealized orbits described by their TLEs by 1--3~km per day.
  \item Model the intercept volume of an laser array of a certain diameter, situated on the Earth, pointing at a given target. 
  \item Calculate the time that the pointing target for the laser array is above the horizon, which equates to the ``total effective beam time.''
  \item Calculate the points of interception between satellites and the beam for a given start time and beaming duration.
  \item Calculate the duration of each intercept and add these intervals together to calculate total intercept time, and hence  the interception time fraction. It is important to note that here, an intercept is defined as the error sphere of the satellites having a non-zero overlap with the uncertainty of the beam.

\end{enumerate}

For each scenario simulation, the configurable inputs of the program are given as follows:

\begin{itemize}
\itemsep=0em
\item Laser array location (defined by latitude \& longitude)
\item Laser pointing target (defined by right ascension \& declination)
\item Starting date and time to energize beam 
\item Duration from initial beam on to beam off (s)
\item Beam diameter (m)
\end{itemize}

\begin{figure}[ht!]
\centering
\includegraphics[width=0.95\linewidth]{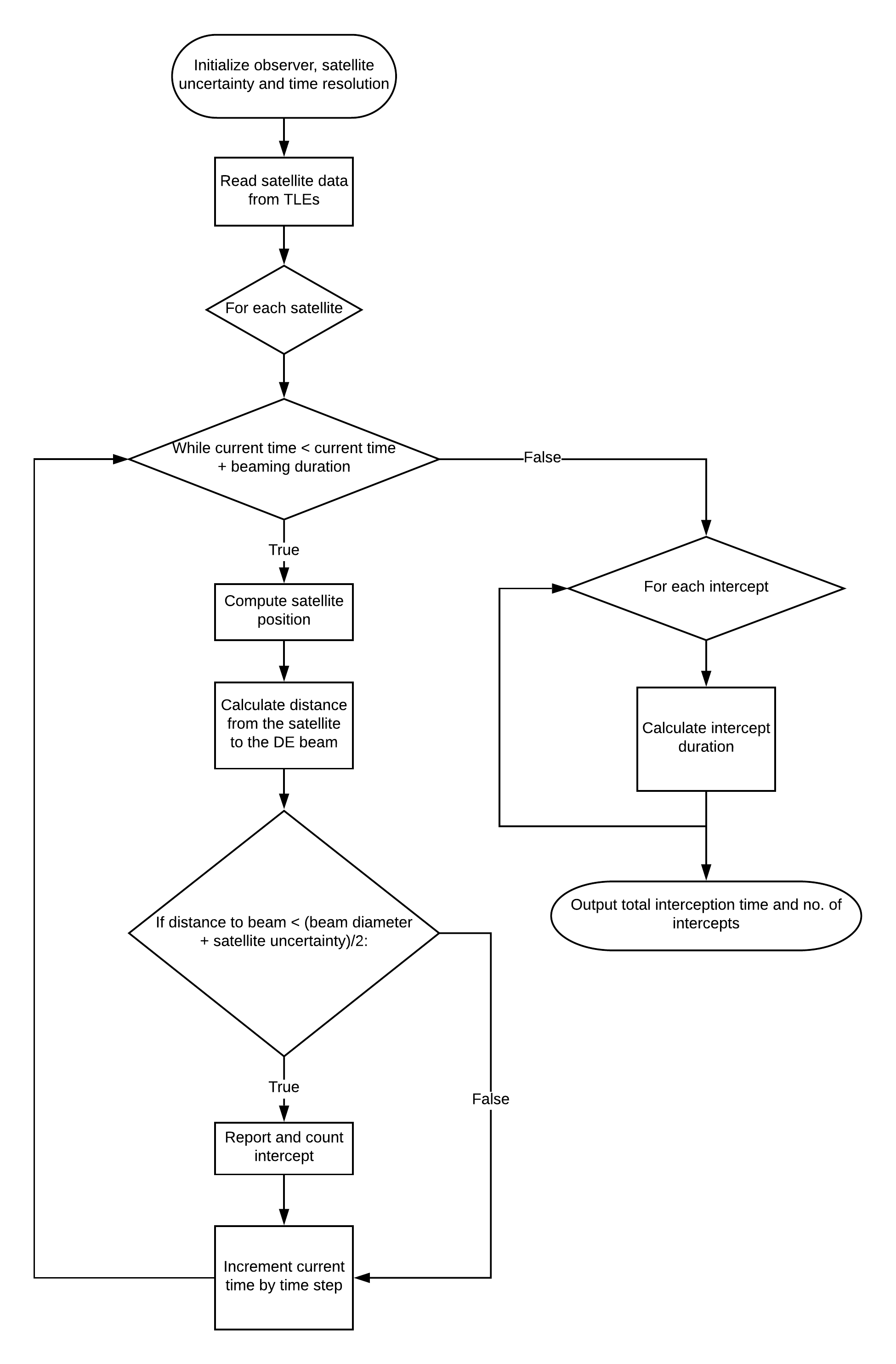}\caption{A flow diagram of Algorithm 1.} 
\label{fig:flow}
\end{figure}

The simulation then generates a list of the intercepts (which contain the position of the particular satellite and the date and time when they intercepted the DE beam) and then returns the total calculated intercept time and number of intercepts for the simulation. The in-depth program is summarized in Algorithm \ref{calcIntTime}, and Figure \ref{fig:flow} gives a flow diagram representation of Algorithm \ref{calcIntTime}.

\begin{algorithm*}
\scriptsize
  \begin{algorithmic}
  \caption{Calculating Number Of Intercepts And Total Intercept Time}
 \label{calcIntTime}
  \State $\mathrm{LEO\_ORBITAL\_SPEED} \gets 7800 \ \mathrm{m \ s}^{-1}$
  \State $\mathrm{satelliteUncertainty} \gets 6000 \ \mathrm{m}$
  \State $\mathrm{resolution} \gets 500$
  \State $\mathrm{stepSize} \gets 1~\mathrm{s}$
  \State $\mathrm{effectiveBeamTime} \gets \Call{$calculateEffectiveBeamTime$}{$startTime, beamingDuration$}$
  \State $\mathrm{totalInterceptTime,numIntercepts} \gets \Call{$calculateTotalInterceptTime$}{$site, target, startTime, beamingDuration, beamDiameter$}$
  \State $\mathrm{interceptFraction} \gets \mathrm{totalInterceptTime / effectiveBeamTime}$  

  \State 
  \Function{calculateTotalInterceptTime}{$\mathrm{site, target, startTime, beamingDuration, beamDiameter}$}
  	\State $\mathrm{satellites}\gets \mathrm{readTLES(TLE)}$
    \State $\mathrm{observer} \gets \mathrm{setup(site,target,startTime)}$
    \State $\mathrm{interceptIntervals} \gets \emptyset$  
    
    \ForAll{$\mathrm{satellite} \in \mathrm{satellites}$} 
    
    	\State $\mathrm{currentTime} \gets \mathrm{startTime}$
        \State $\mathrm{intercepts} \gets \emptyset$
        
        \State $\mathrm{satellitePosition} \gets \Call{$computePosition$}{$observer, target$}$
        \State $\mathrm{distanceToBeam} \gets \Call{$calculateDistance$}{$satellitePosition$}^{\dagger}$
 
        \State
        
        \While{$ \mathrm{currentTime} < (\mathrm{currentTime} + \mathrm{beamingDuration})$}
        
            \If{$ \mathrm{distanceToBeam} < (\mathrm{beamDiameter} + \mathrm{satelliteUncertainty})/2$} 
    
                \State $\mathrm{intercept} \gets \mathrm{currentTime}$
                \State $\mathrm{intercepts.add(intercept)}$
                \State $\mathrm{numIntercepts} \gets \mathrm{numIntercepts + 1}$
    
            \EndIf
            
            \State $\mathrm{timestep} \gets \mathrm{distanceToBeam / (resolution \times LEO\_ORBITAL\_SPEED)} $
            
            \State $\mathrm{currentTime} \gets \mathrm{currentTime + timestep}$
        \EndWhile
        \State 
        
        \If{$\exists\ \mathrm{intercepts}$}
        	\State $\mathrm{interceptInterval} \gets \mathrm{(intercepts[0], intercepts[intercepts.length-1])}$
            \State $\mathrm{interceptIntervals.add(interval)}$
        \EndIf
    \EndFor
    \State 
    \ForAll {$\mathrm{interval} \in \mathrm{interceptIntervals}$}
    	\State $\mathrm{totalInterceptTime} \gets \mathrm{totalInterceptTime + interval}$
    \EndFor
    \State 
          \State \Return $\mathrm{totalInterceptTime, numIntercepts}$
\EndFunction
\end{algorithmic}
\end{algorithm*}

%   \begin{algorithm*}
%   \begin{algorithmic}
%   \scriptsize
%     \caption{Calculating Effective Beam Time}
%  \label{calcEffTime}
%  \Function{calculateEffectiveBeamTime}{$\mathrm{startTime,beamingDuration}$}
%  	\State $\mathrm{timeAboveHorizon} \gets 0$
%     \State $\mathrm{timeAbove30Degrees} \gets 0$
%     \State $\mathrm{currentTime} \gets \mathrm{startTime}$
%     \State $\mathrm{endDateTime} \gets \mathrm{startTime + beamingDuration}$
%     \State
%     \While{$\mathrm{currentTime < endDateTime}$}
%   		\State $\mathrm{pointingAltitude} \gets \Call{$computePosition$}{$observer, target$}$
        
%         \If {$\mathrm{pointingAltitude} \in \lbrack0,90\rbrack$}
%         	 \State $\mathrm{timeAboveHorizon} \gets \mathrm{timeAboveHorizon+stepSize}$
%         \EndIf 
                     
%         \If {$\mathrm{pointingAltitude} \in \lbrack30,90\rbrack$}
%         	  \State $\mathrm{timeAbove30Degrees} \gets \mathrm{timeAbove30Degrees+stepSize}$
%         \EndIf 
%         \State $\mathrm{currentTime} \gets \mathrm{currentTime + stepSize}$
%     \EndWhile
%     \State
%     \State \Return $\mathrm{timeAboveHorizon, \ timeAbove30Degrees}$
% \EndFunction
%   \end{algorithmic}
%   \end{algorithm*}

 \subsubsection{Satellite distance calculation}
 \label{Satellite tolerance and distance calculation}

\begin{figure}[ht!]
\centering
\includegraphics[width=1\linewidth]{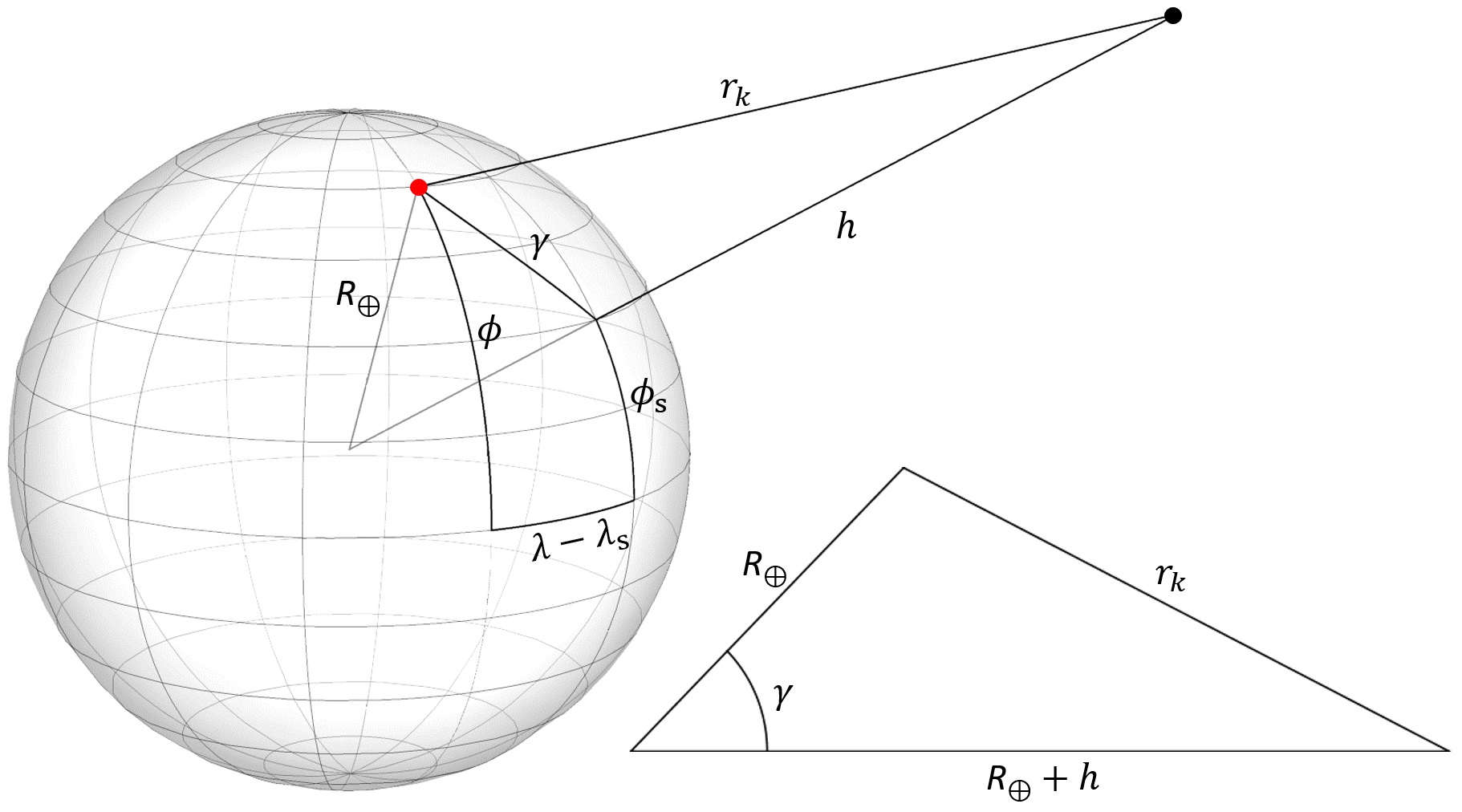} 
\caption{Left: A model of the Earth showing the locations of a laser array (red circle), a satellite (black circle), their coordinates, and the angle $\gamma$ between them. $r_k$ is the distance between the laser and satellite. \newline Right: The law of cosines is used to calculate $r_k$ using $R_{\oplus}, h,$ and $\gamma$.}
\label{fig:sat_tol}
\end{figure}

The distance from the center of the beam to the expected position of the satellite, which we will call $p$, is calculated using a function (\texttt{calculateDistance()}) as follows. The user inputs the laser's latitude $\phi$, the laser's longitude $\lambda$, and the target's right ascension and declination which are then converted to its altitude angle $\alpha$ and azimuth angle $\beta$. \texttt{PyEphem} uses $\phi$ and $\lambda$ to return the latitude $\phi_\mathrm{s}$, longitude $\lambda_\mathrm{s}$, altitude angle $\alpha_\mathrm{s}$, azimuth angle $\beta_\mathrm{s}$ and elevation 
$h$ above sea level of the satellite. First, the distance $r_k$ from the laser array site to the satellite is calculated by the law of cosines:

\begin{equation}
r_k = (R_{\oplus}+h)[1+(\textstyle\frac{R_{\oplus}}{R_{\oplus}+h})^2 - 2(\frac{R_{\oplus}}{R_{\oplus}+h}) \cos{\gamma}⁡]^{1/2},
\end{equation}
where $R_{\oplus}$ is the radius of Earth and $\gamma$ is the angle between the laser and satellite's radius vectors shown in Fig. \ref{fig:sat_tol}. From the dot product in spherical coordinates, we have                                    

\begin{equation}
\cos{\gamma}⁡=\cos⁡{\phi_\mathrm{s}} \cos⁡{\phi }\cos⁡{(\lambda-\lambda_\mathrm{s})}+\sin⁡{\phi_\mathrm{s}} \sin⁡{\phi}.
\label{eqn:14}
\end{equation}

\begin{figure}[ht!]
\centering
\includegraphics[width=1.1\linewidth]{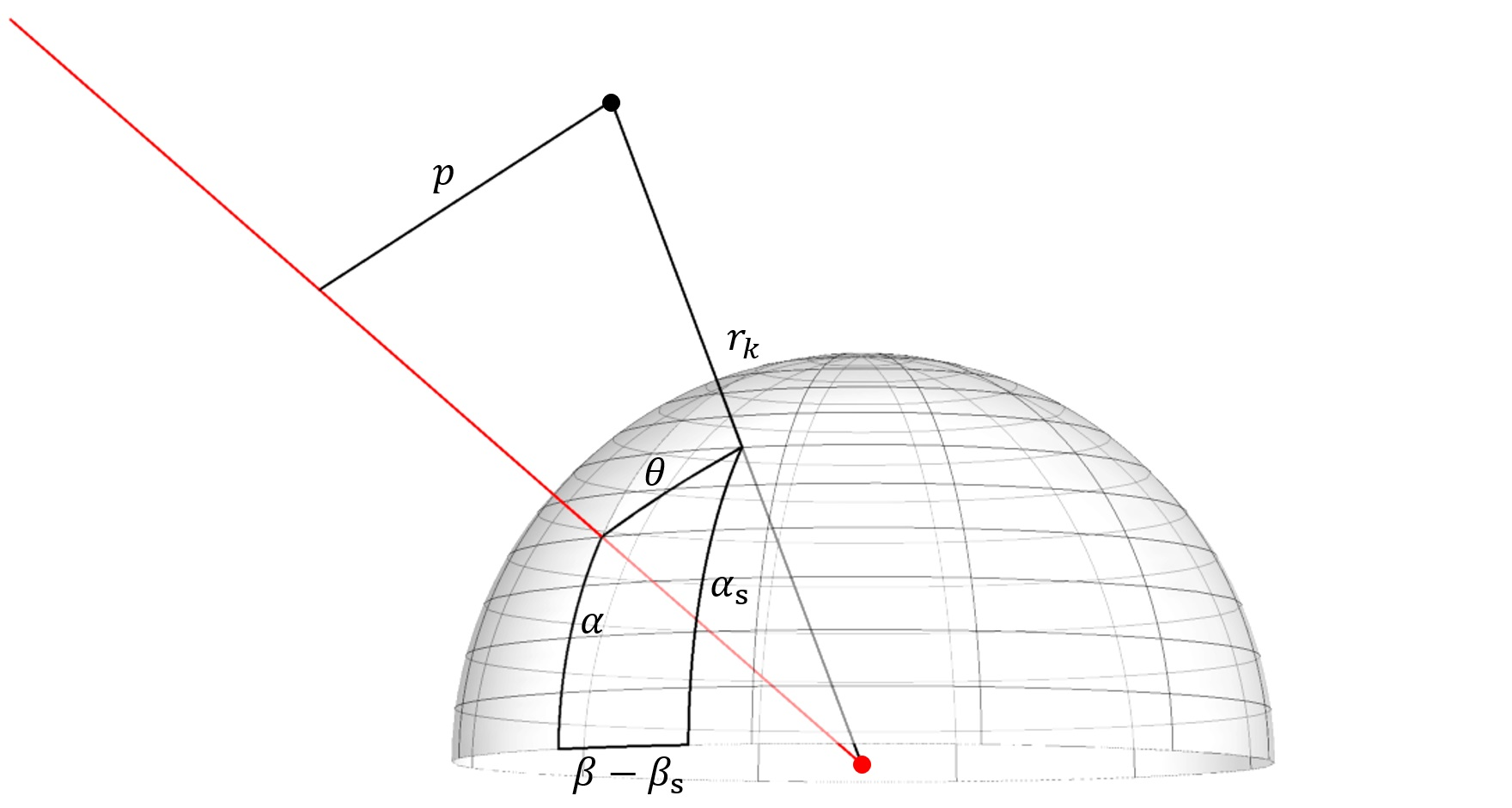} 
\caption{A celestial sphere centered at the location of the laser array (red circle), showing the altitude angle $\alpha$ and azimuth angle $\beta$ of the beam, as well a satellite (black circle) with altitude angle $\alpha_\mathrm{s}$ and azimuth angle $\beta_\mathrm{s}$. $\theta$ is the angular displacement between the laser and satellite, $r_k$ is the distance from the laser to the satellite, and $p$ is the shortest distance from the satellite to the center of the beam.} 
\label{fig:sat_tol2}
\end{figure}

Next, the angle $\theta$ between the beam and the satellite (shown in Fig. \ref{fig:sat_tol2}) is used with $r_k$ to calculate $p$. The computation of $\theta$ is similar to that of $\gamma$:

\begin{eqnarray}
\cos{\theta} &=& \cos⁡{\alpha_\mathrm{s}} \cos⁡{\alpha }\cos⁡{(\beta-\beta_\mathrm{s})}+\sin⁡{\alpha_\mathrm{s}} \sin⁡{\alpha} \\ \\
\sin{\theta} &=& (1-\cos{\theta}^2)^{1/2}.
\end{eqnarray}

Finally, the distance from the center of the beam to the expected position of the satellite is

\begin{equation}
p = r_k \sin{\theta}.
\end{equation}
If $p$ is within $(d+\delta_\mathrm{sat})/2$, an intercept is counted. We neglect the size of the satellite and the laser half-angle for the reasons described in Sec. \ref{Intercept probability}.

\subsubsection{Time step error analysis}
\label{Time step error analysis}

In computational simulations, there is naturally an error in output values due to the finite resolution of their calculations. In our case, this error originates from the size of the time step between data points. Our simulation uses a dynamic time step $\Delta t$ equal to the distance $p$ between the expected position of the satellite and the center of the beam divided by the speed of low-Earth orbital satellites (denoted as $v_\mathrm{LEO} = 7800$~m/s) and scaled by a resolution number $N_\mathrm{res}$:

\begin{equation}
\Delta t \equiv \frac{p}{N_\mathrm{res} v_\mathrm{LEO}}.
\end{equation}

\begin{figure}[ht!]
\centering
\includegraphics[width=1\linewidth]{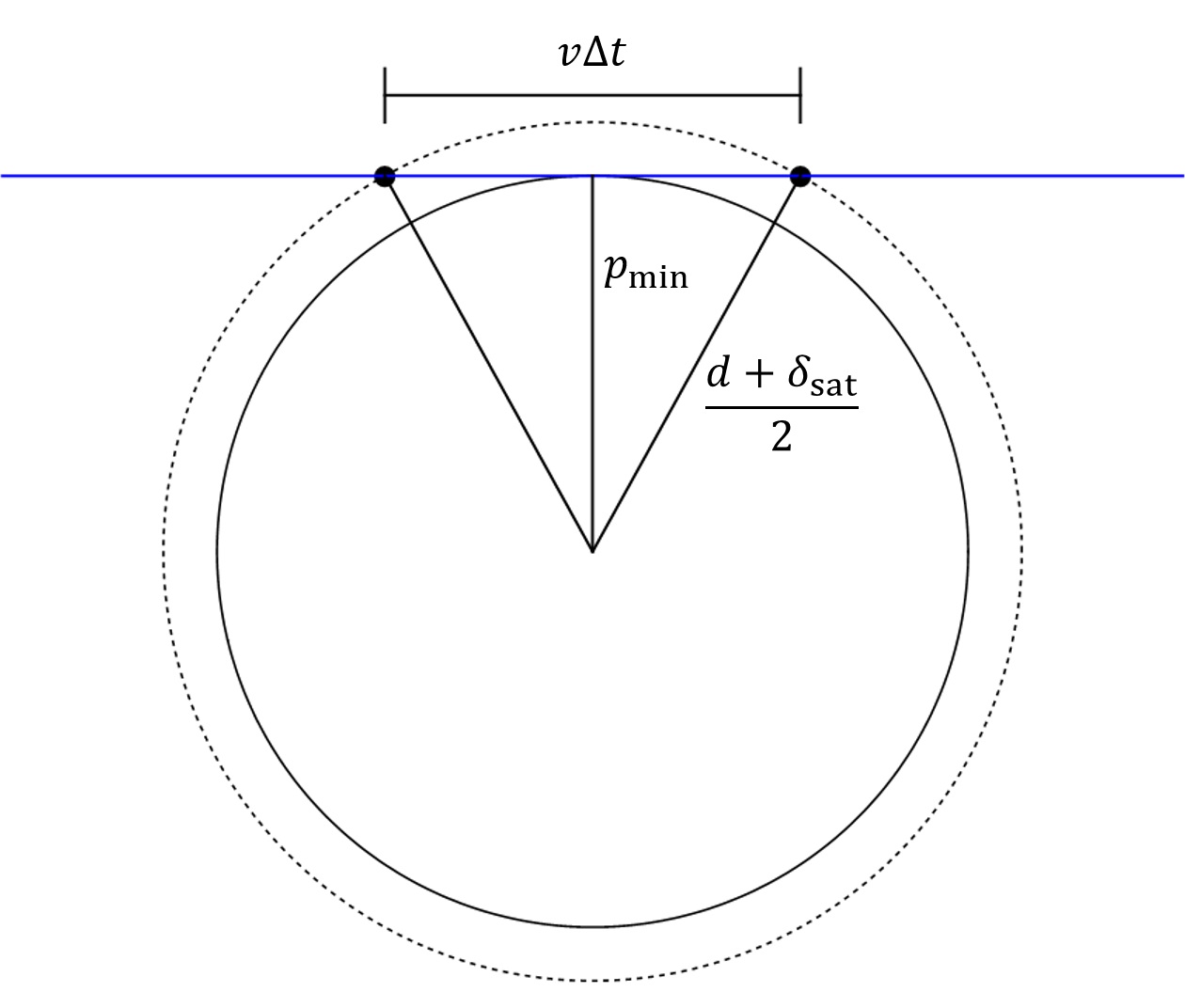} 
\caption{The intercept area as seen by the simulation (solid circle) is dependent on the time resolution. The dotted circle is the true intercept area shown in Fig. \ref{fig:hitbox}. The blue line represents a satellite's trajectory as described by its TLE file, and the black dots represent two consecutive calculations of its location separated by the time interval $\Delta t$. In this case, an intercept is not counted, even though the expected position of the satellite passes through the intercept area. Thus, the effective intercept area is that of a circle with a diameter $2p_\mathrm{min}$ ($p_\mathrm{min}$ is the closest distance from the satellite's expected trajectory to the center of the beam) rather than $d+\delta_\mathrm{sat}$.}
\label{fig:error}
\end{figure}

The time resolution only plays a role when a satellite enters and exits the beam. The most extreme error in the measurement of intercept time that can occur is when the data points for the satellite's position are at the very edge of the intercept area. In this case,

\begin{equation}
%\begin{split}
p = \frac{d+\delta_\mathrm{sat}}{2}
%\end{split}
\end{equation}

\begin{equation}
%\begin{split}
\label{eqn: delta t}
\Delta t = \frac{d+\delta_\mathrm{sat}}{2 N_\mathrm{res} v_\mathrm{LEO}}.
%\end{split}
\end{equation}
The error $\delta t_\mathrm{int}$ is

\begin{equation}
%\begin{split}
\delta t_\mathrm{int} = 2\Delta t = \frac{d+\delta_\mathrm{sat}}{N_\mathrm{res} v_\mathrm{LEO}}.
%\end{split}
\end{equation}
The total error in intercept duration for $N_\mathrm{int}$ number of intercepts is then

\begin{equation}
%\begin{split}
\label{eqn: total delta t}
\delta T_\mathrm{int} = \sum \delta t_\mathrm{int} = N_\mathrm{int}\frac{d+\delta_\mathrm{sat}}{N_\mathrm{res} v_\mathrm{LEO}}.
%\end{split}
\end{equation}

\begin{figure}[ht!]
\centering
\includegraphics[width=1\linewidth]{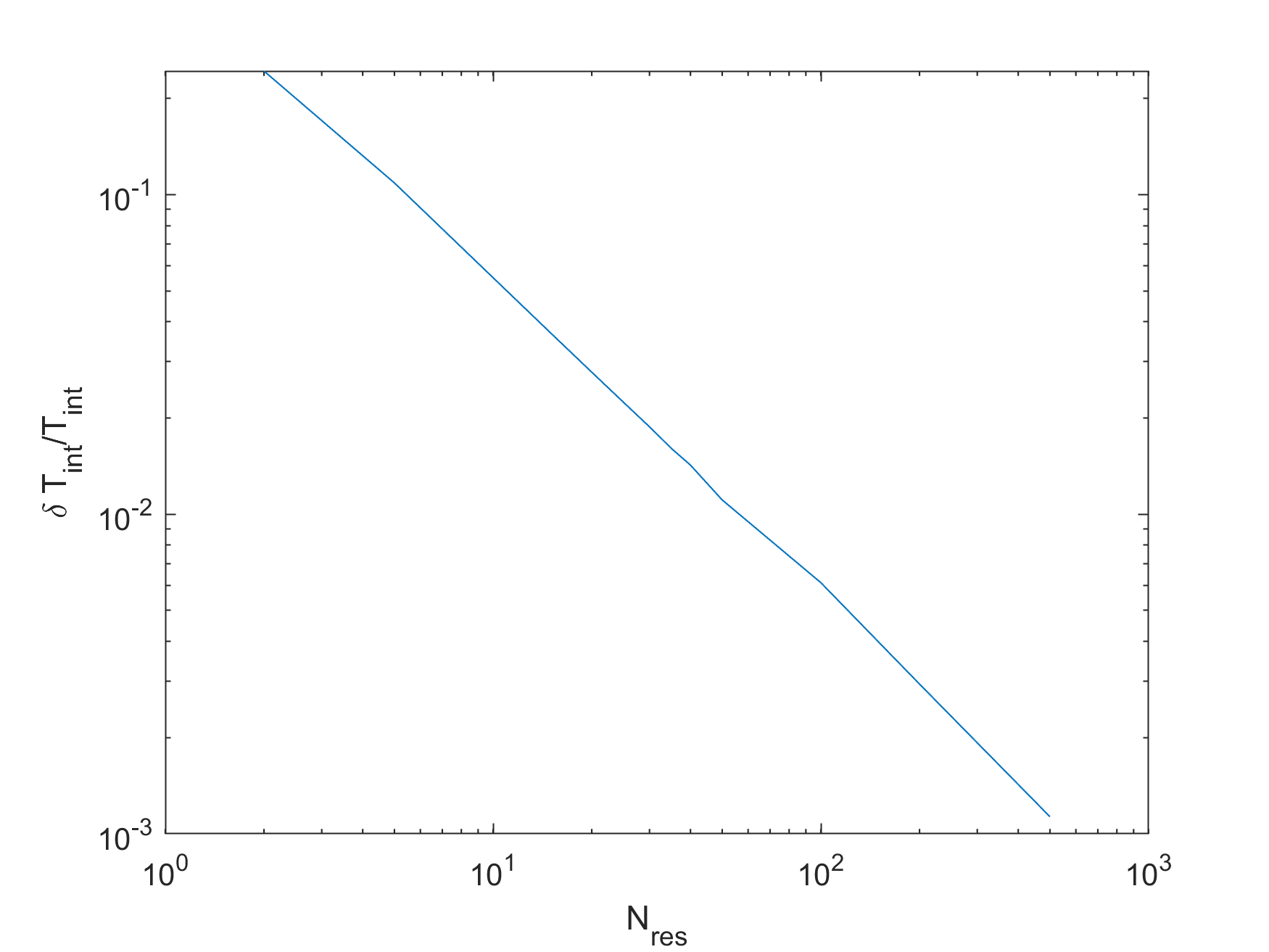} 
\caption{Fractional error of intercept duration as a function of resolution number for a simulation of a 90 day mission. The curve closely follows the $N_\mathrm{res}^{-1}$ relationship predicted by Eq. (\ref{eqn: total delta t}). The true value of $T_\mathrm{int}$ (total intercept duration) was interpolated using the simulated data points and Eq. (\ref{eqn: total delta t}), with a factor of two introduced to account for averaging. For our choice of $N_\mathrm{res} = 500$, the error is about 0.1\%, which is sufficiently low for our purposes.}
\label{fig:convergence}
\end{figure}

The time resolution also produces an error in $N_\mathrm{int}$, which we will call $\delta N_\mathrm{int}$. The intercept time for some satellites may be less than $\Delta t$ if their error sphere surface intersects the edge of the beam, in which case the simulation will not report an intercept. In the most extreme case, two data points are collected at $p = (d + \delta_\mathrm{sat})/2$ before and after the expected trajectory of the satellite crosses the intercept area (see Fig. \ref{fig:error}). This condition creates a new effective intercept area equal to the area of a circle with diameter $2p_\mathrm{min}$ rather than $d + \delta_\mathrm{sat}$, where $p_\mathrm{min}$ is the closest distance from the satellite's expected trajectory to the center of the beam. Using the satellite's speed $v$ and assuming the beam is stationary (which it will be on average), we have

\begin{eqnarray}
p_\mathrm{min} &=& \frac{1}{2}\sqrt{(d+\delta_\mathrm{sat})^2-(v \Delta t)^2} \\
A_\mathrm{Intercept} &=& \frac{\pi}{4}[(d+\delta_\mathrm{sat})^2-(v \Delta t)^2].
\end{eqnarray}
The fractional change in $n_\mathrm{int}$ intercepts with a single satellite, which is equal to the fractional change in intercept area, is then

\begin{eqnarray}
\frac{n_\mathrm{int}-\delta n_\mathrm{int}}{n_\mathrm{int}} &=& \frac{(d+\delta_\mathrm{sat})^2-(v \Delta t)^2}{(d +\delta_\mathrm{sat})^2} \\
\therefore \delta n_\mathrm{int} &=& n_\mathrm{int} \frac{(v \Delta t)^2}{(d+\delta_\mathrm{sat})^2},
\end{eqnarray}
where $\delta n_\mathrm{int}$ in the error in the number of intercepts with a single satellite. Using Eq. (\ref{eqn: delta t}) to replace $\Delta t$, this error is

\begin{equation}
%\begin{split}
\delta n_\mathrm{int} = n_\mathrm{int} \frac{v^2}{(2 N_\mathrm{res} v_\mathrm{LEO})^2}.
%\end{split}
\end{equation}
For all satellites, we have

\begin{eqnarray}
\delta N_\mathrm{int} &=& \sum n_\mathrm{int} \frac{v^2}{(2 N_\mathrm{res} v_\mathrm{LEO})^2} \\
&\leq& \sum n_\mathrm{int} \frac{v_\mathrm{LEO}^2}{(2 N_\mathrm{res} v_\mathrm{LEO})^2}\\
&=& \frac{N_\mathrm{int}}{4 N_\mathrm{res}^2}.
\end{eqnarray}
The above quantity is the highest expected error in number of intercepts with all satellites for a given simulation. As $N_\mathrm{res}$ increases, the simulation outputs become more accurate, but they consume greater computational resources. The value for $N_\mathrm{res}$ was chosen such that it was in a regime where our results were invariant as a function of the time step size, to the reported accuracy. Our choice $N_\mathrm{res} = 500$ is safely in this regime without being unnecessarily large (see Fig. \ref{fig:convergence}). For simulations with $d = 10$~km and $\delta_\mathrm{sat} = 6$~km (worst case) this yields $\delta t_\mathrm{int} = 2$~ms. Due to their dependencies on $N_\mathrm{res}$, the results for number of intercepts converged much more rapidly than those for intercept duration, so in general $\delta N_\mathrm{int} << 1$.

\subsubsection{Calculating distances from beam to all satellites at a given instant}
\label{Calculating distance from satellites to beam}

As another heuristic and useful tool to determine the probability of intercepting Earth-orbiting satellites, calculating the shortest distance from each satellite to the beam at a given distance was implemented. This was carried out for validation purposes as well as to give a clear idea of the distribution of distances of satellites from the beam for any given time, place and target.
  
% \begin{algorithm*}
% \scriptsize
% \caption{Calculate distance from beam to all satellites simulator}
%  \label{calcDist}
%   \begin{algorithmic}
%   \Function{calculateDistanceSnapshot}{$\mathrm{site, target, time}$}
%   	\State $\mathrm{satellites}\gets \mathrm{readTLEFile(TLEFile)}$
%     \State $\mathrm{observer} \gets \mathrm{setup(site,target,time)}$ 
%     \State $\mathrm{distances} \gets \emptyset$  
%     \State
%     \ForAll{$\mathrm{satellite} \in \mathrm{satellites}$} 
%     	\State $\mathrm{currentTime} \gets \mathrm{time}$        
%         \State $\mathrm{satellitePosition} \gets \Call{$computePosition$}{$observer, target$}$
%         \If{$\mathrm{satellitePosition.altitude > 0}$} 
%         	\State $\mathrm{distanceToBeam} \gets \Call{$calculateDistance$}{$satellitePosition,observer$}^{\dagger}$
%             \State $\mathrm{distances.add(distanceToBeam)}$
%         \EndIf
        
%     \EndFor
%     \State
%   \State \Return $\mathrm{distances}$
% \EndFunction
% \end{algorithmic}
%   \end{algorithm*}

The shortest distance between the beam and a given satellite is calculated in Sec. \ref{Satellite tolerance and distance calculation}. Refer to Fig. \ref{fig:sat_tol2} for a geometric visualization. This calculation yields the distance from the expected position of the satellite to the center of the beam, not the worst-case minimum distance from the satellite to the edge of the beam. In other words, this distance should not be less than $(d+\delta_\mathrm{sat})/2$ before the DE is gated off. Satellites that are under the horizon ($\alpha_\mathrm{s }< 0^\circ$) or behind the laser array ($\theta > 90^\circ$) are ignored.

\subsubsection{Scenario planning}
\label{Scenario Planning}

In order to answer the research questions, a number of laser array base sites were chosen for the potential placement of a multi-element laser array that will be used in future missions. These sites, given in Table \ref{tab:sites}, offer preferable conditions for observational astronomy due to their high altitudes and low air turbulence, which also makes them good candidates for DE base sites. The targets chosen pertain to specific future proposed NASA missions and are tabulated in Table \ref{tab:targets}.\\

\begin{table*}[ht!]
\centering
\scriptsize
\caption{Laser array base sites.}\label{tab:sites}
 \begin{tabular}{c c c c c c}
 \hline\hline
Laser Array Base Site & Location & Latitude (deg) & Longitude (deg) & Declin. range (deg) & Altitude (m) \\ 
\hline
 Barcroft Station & Mono County, California & 37.584 & -118.237  & [-22.417,97.584] & 3100 \\ 

ALMA & Atacama Desert, Santiago, Chile & -23.023 & -67.755 & [-83.023,36.977] & 3158 \\ 
Haleakala Observatory & Kula, Hawaii & 20.710 & -156.253 & [-39.290,80.710] & 3052 \\ 

 South Pole & Antarctica & -90.000 & 0.000 & [-30.000,60.000] & 2800 \\ 
  \hline
\end{tabular}
\end{table*}

The laser array continuously tracks the specified target for the duration of the simulation, but can only activate when the target is above the horizon. It is therefore important for any given mission to consider both the time when the target is above the horizon, as well as the intercept fraction.\\

The first scenario that we are exploring through our simulations is the long-term interstellar mission to the Alpha Centauri system. Proxima Centauri is the nearest solar system to our own, and is approximately 4.24 light years away. Proxima Centauri has at least one confirmed exoplanet (Proxima b), and based on Kepler data, the number of planets per star is approximately unity \citep{kepler}. Interstellar travel is exceedingly difficult and presently not feasible for humans to undertake. However, we can instead use ``remote sensing'' through lightweight electronic systems (i.e., wafer spacecraft) to allow for exploration across vast distances \citep{pml2016}.\\

In the mission proposed by \citet{pml2016}, it is shown through analysis that a 100~GW DE system can accelerate a wafer spacecraft that weighs 1 gram to 0.25c in a few minutes of laser illumination, reaching the Alpha Centauri region within about 20 years. These spacecraft will likely need optical beacons on board for tracking and DE phasing purposes. This short acceleration time would potentially allow for hundreds of missions per day, or approximately one-hundred thousand missions per year. The ability to propel a multitude of spacecraft in large numbers means spreading the risk of failed missions over the constellation of spacecraft. This method of exploration can be reasoned about in a similar vein to the concept of an r-strategist organism in ecology. In ecology, an r-strategist is an opportunist that produces a large number of inexpensive offspring, which works very well in surviving in unknown and unstable environments \citep{rstrategist}. In our case, we are sending many spacecraft out, and can tolerate single spacecraft failures due to the large swarm strategy we would use. \\

Wafer-scale spacecraft are not the only type of payload to be explored. Heavier payloads can be accelerated through the same DE system, for rapid transit to closer targets such as the Moon, Mars, and Pluto. For example, in these regimes, a 100~kg payload may be accelerated to 0.01c, and a 10-ton payload potentially to over 1000~km/s. Since fuel is no longer an issue, and the directed energy propulsion system is on Earth, such missions would become more cost-effective and rapid to deploy. For such missions in the future, it is possible to decelerate the spacecraft by utilizing either a remotely controlled mirror to reflect the directed energy beam from Earth back at the spacecraft, or an equivalent DE system located at the destination.\\

Of particular importance in the shorter-term vision of Starlight are missions that involve using a DE beam to power high performance ion engines on the spacecraft, allowing high mass interplanetary missions at modest speeds \citep{pml2016}. These are not for relativistic flight applications but do allow much higher mass missions in the solar system. High mass tugs back and forth to the Moon or Mars are one example of this. \\

\begin{table}[ht!]
\centering
\caption{Targets.}\label{tab:targets}
 \begin{tabular}{c c}
 \hline\hline
Target & Distance from Earth \\ 
 \hline
 Alpha Centauri A & 4.367 ly \\ 
 
 The Moon & 384,400 km \\ 
 
 Mars & 54.6 million km \\ 
 
 Pluto & 7.5 billion km \\ 
 \hline
\end{tabular}
\end{table}

Appropriate beaming durations and beam diameters were chosen for each simulation depending on the proposed scenario. For example, 1000 seconds was chosen for the Alpha Centauri target scenario, as this is a conservative upper-bound on the time that the directed energy must be energized in order to accelerate gram-scale wafer-crafts to relativistic speeds. Other time periods include a 3-month duration for Mars, a 1-year period for the moon, and a 3-year period for Pluto. A 100~m beam diameter was chosen for the moon-based mission, whereas a 10~km diameter was chosen for all other scenarios. 

Each mission scenario was simulated for each of the four candidate laser array base sites to investigate the relative importance of the location of the laser array for any given mission. Simulations were run on a Desktop computer, with an Intel Core 2 Quad Q6600 CPU @ 2.40GHz, 8GB of RAM, and Windows 7 Ultimate edition.

\section{Results} 
\label{Results}

With the methodology established, we are subsequently able to carry out the simulations and measure a number of interesting results. For each simulation, the following results were calculated:

\begin{itemize}

\item \textit{t desired} or the desired illumination time denotes the time we wish to keep the DE beam on from the starting date. 
\item \textit{t above 0$^{\circ}$} denotes the total time that the desired target is above the horizon (the total time we can aim the DE beam at the target without going through the Earth). 
\item \textit{t above 30$^{\circ}$ }is a similar metric as above, except now only measures the time when the beam is 30$^{\circ}$ above the horizon when pointing at the target. At less than 30$^{\circ}$ above the horizon, the beam is usable, but atmospheric absorption and scattering of the DE become more significant and reduce the intensity at the target.
\item \textit{No. Int.s} represents the number of times that satellites have intercepted the beam during the mission simulation. 
\item \textit{Int. duration} represents the total time that satellites have intercepted the beam during the simulation. It is important to note that we have ensured in our simulation (using an intercept interval merging algorithm), that two or more satellites intercepting the beam at the same time are not double-counted. 
\item The \textit{Int. Fraction} represents the fraction of total time that satellites intercept the beam out of the time that the desired target is above the horizon (\textit{Int. duration}/\textit{t above 0$^{\circ}$}).
\end{itemize}

The start date and time of each simulation was set to March 20th, 2018 at midnight. The error bounds were set by use of the equations in Sec. \ref{Time step error analysis}.

\subsection{Alpha Centauri}

Simulations were run for a low mass ($\approx 10$~g) wafer scale spacecraft interstellar mission, targeting Alpha Centauri, using a beaming duration of 1000 seconds, a beam diameter set at 10 km (for a 100~GW DE beam), and only considering active satellites.

In general, the simulated interstellar cases with a beaming duration of 1000 seconds or less did not yield an intercept. This is consistent with our analytic solution, as the number of intercepts per second was calculated at 0.000232 $\mathrm{s}^{-1}$, which corresponds to, on average, 0.232 intercepts in 1000 seconds.\\

From the lack of intercepts observed in simulation and the statistical argument from Sec. \ref{Satellite beam crossing duration and frequency}, we conclude that at any given location on Earth, there should always exist a window of at least 1000 seconds where the beam never intercepts any satellites. Therefore, the ability to launch these wafer satellites without the DE beam intercepting another active satellite (thus eliminating the need to gate the DE beam off during the acceleration period) is feasible.

\subsection{The Moon}

Simulations pertaining to missions to the the moon were conducted. These would be done primarily using much lower power ($<$100 MW) driving high performance ion engines. Since such missions are much closer and could be frequent, we use a beam diameter of 100 m and allow for continuous operation for a  year and only consider active satellites. The results are tabulated in Table \ref{tab:moon}.

For the long-term lunar operations, it is critical that the transmission proportion be very high. This means that the beam should not be gated off for extended periods of time. As we can see from Table \ref{tab:moon}, this is overwhelmingly the case, with an intercept fraction ranging from just $6.06 \times 10^{-5}$ to $2.66 \times 10^{-4}$. Note that the intercept fraction of the moon mission is an order of magnitude lower than that of the Mars and Pluto missions because of the decreased beam size (100~m rather than 10~km).

\begin{table*}[ht!]
\centering
\scriptsize
\caption{Simulation: Moon.}\label{tab:moon}
 \begin{tabular}{c c c c c c c c}
 \hline\hline
Laser Array Base Site & \textit{t} desired & \textit{t} above 0$^{\circ}$ & \textit{t} above 30$^{\circ}$ & No. Int.s & Int. duration & Int. fraction\\ 
 \hline
 Barcroft Station & 1 year & 180.338 days & 89.423 days & 1528 $\pm <$ 1 & 19.1 $\pm < 0.1$ mins & $ 7.34 \times 10^{-5}$ \\
 \hline
 Atacama & 1 year & 181.914 days & 110.780 days & 1358 $\pm < 1$ & 15.9 $\pm < 0.1$ mins & $6.06 \times 10^{-5}$ \\ 
 \hline
 Haleakala & 1 year & 181.114 days & 111.266 days & 1427 $\pm < 1$ & 16.8 $\pm < 0.1$ mins & $6.43 \times 10^{-5}$ \\ 
 \hline
 South Pole & 1 year & 186.794 days & 0 days & 4737 $\pm < 1$ & 71.6 $\pm \ 0.1$ mins & $2.66 \times 10^{-4}$ \\ 
 \hline
\end{tabular}
\end{table*}

\begin{table*}[ht!]
\centering
\scriptsize
\caption{A sample distance snapshot simulation.}\label{tab:moonSnapshot}
 \begin{tabular}{c c c c c c}
 \hline\hline
Satellite name & Distance (km) & Angular Disp. (deg) & Altitude (deg) & Azimuth (deg) & Elevation (km) \\ 
 \hline
 AEROCUBE 7C & 78 & 9.03399 & 67.388 & 190.662 & 453 \\ 
 
 YAOGAN 4 & 225&	15.65786&	50.572&	158.021	& 654 \\
 
 AEROCUBE 7B &303&	36.98796&	63.477&	263.674 &	453 \\

GOMX 1 & 334&	20.3515	&39.14	&180.285	&636	\\				

YAOGAN 29& 522&	47.83306&	61.916&	292.552&	632\\					

GLOBALSTAR M079& 529&21.60443&	79.771&	205.448&	1416\\					

APRIZESAT 5 &626	&30.04942&	29.449	&179.673&	688	\\				

IRIDIUM 136 &1286&	87.71908 &32.554&352.763&	785	\\				

SUSAT &1306&	57.07354&	7.509&	211.847&	374	\\				

CYGFM05    &1314&	61.79961 &14.582&122.095&	522	\\				

  \vdots &  \vdots	&  \vdots &  \vdots &  \vdots &	 \vdots \\

 SPEKTR R	&332732	&86.7313	&2.537&	92.876	&327217 \\
 \hline
\end{tabular}
\end{table*}

Table \ref{tab:moonSnapshot} represents a demonstrative output of a distance snapshot simulation, beaming from Barcroft to the Moon (Alt = 59.489$^\circ$, Az = 180.742$^\circ$), at 2018/02/01 midnight. The table (truncated for brevity) shows every active satellite, sorted by distance from the laser array. The distances reported are the distances from the center of the beam to the positions of satellites as calculated from their TLEs, not the worst-case minimum distances from the satellites to the edge of the beam. Using such simulations can be useful in determining optimal times to decide mission times and sites.

\subsection{Mars}

In addition, for missions involving beaming to Martian based systems, simulations were made with a beaming duration of 3 months, with a beam diameter set at 10~km, and only considering active satellites. The results are tabulated in Table \ref{tab:mars}.

\begin{table*}[ht!]
\centering
\scriptsize
\caption{Simulation: Mars.}\label{tab:mars}
 \begin{tabular}{c c c c c c c c}
 \hline\hline
Laser Array Base Site & \textit{t} desired & \textit{t} above 0$^{\circ}$ & \textit{t} above 30$^{\circ}$ & No. Int.s & Int. duration & Int. Fraction \\ 
 \hline
 Barcroft Station & 3 months & 36.00 days & 1.89 days & 535 $\pm < 1$ & 21.1 $\pm < 0.1$ mins & $4.07 \times 10^{-4}$ \\ 

 Atacama & 3 months & 50.35 days & 32.81 days & 1207 $\pm < 1$ & 37.1 $\pm < 0.1$ mins & $5.11 \times 10^{-4}$ \\ 

 Haleakala & 3 months & 40.74 days & 21.27 days & 688 $\pm < 1$ & 23.7 $\pm < 0.1$ mins & $4.04 \times 10^{-4}$ \\ 

 South Pole & 3 months & 90.00 days & 0 days & 26284 $\pm < 1$ & 13.5 $\pm < 0.1$ hours & $6.24 \times 10^{-3}$ \\ 
\hline
\end{tabular}
\end{table*}

\subsection{Pluto}

Lastly, simulations for long-term DE mission scenarios to Pluto were also conducted. For these missions the beaming duration was set at 3 years, with a beam diameter set at 10~km, and only considering active satellites. The results are tabulated in Table \ref{tab:pluto}.\\

\begin{table*}[ht!]
\centering
\scriptsize
\caption{Simulation: Pluto.}\label{tab:pluto}
 \begin{tabular}{c c c c c c c c}
 \hline\hline
Laser Array Base Site & \textit{t} desired & \textit{t} above 0$^{\circ}$ & \textit{t} above 30$^{\circ}$ & No. Int.s & Int. duration & Int. Fraction \\ 
 \hline
 Barcroft Station & 3 years & 441.63 days & 37.75 days & 3296 $\pm < 1$ & 2.06 $\pm < 0.01$ hours & $1.94 \times 10^{-4}$ \\ 

 Atacama & 3 years & 611.98 days & 398.98 days & 4085 $\pm < 1$ & 2.06 $\pm < 0.01$ hours & $1.40 \times 10^{-4}$ \\ 

 Haleakala & 3 years & 497.78 days & 262.33 days & 3079 $\pm < 1$ & 1.72 $\pm < 0.01$ hours& $1.44 \times 10^{-4}$ \\ 

  South Pole & 3 years & 1095.00 days & 0 days & 45200 $\pm < 1$ & 24.8 $\pm < 0.1$ hours & $9.44 \times 10^{-4}$ \\ 
 \hline
\end{tabular}
\end{table*}

As we can observe in Tabs. \ref{tab:moon}, \ref{tab:mars}, and \ref{tab:pluto}, the probability of any active satellite intercepting with the beam is small.

\subsection{Beam Intercept Fraction}

It was deemed appropriate to find the relationship between the total beaming duration and the proportion of time that the beam needed to be gated off. This allows us to establish whether the simulated data was consistent with our mathematical analysis.\\

% \begin{figure}[ht!]
% \centering
% \includegraphics[width=.8\linewidth]{Beam_Duration_and_Intercept_Fraction-new_sims-d}  
% \caption{DE beam intercept fraction plotted against total beaming time for a different combinations of launch site and DE pointing target. The excess observed for the shorter durations is simply due to the combination of initial conditions and input arguments for the simulation. These included the initial beaming date, launch site, and target. Simulations performed at the South Pole exhibited higher incidences of satellite interception, due to intercepting more LEO satellites likely because of a horizon projection effect.}
% \label{fig:BDVI}
% \end{figure}

As the beaming duration increases, the beam intercept fraction for most simulations tends to approach the theoretical fraction we have calculated in Sec. \ref{Analytical Approximation}.  Each date was chosen such that the target would be within in the horizon of the beam, and as close to the epoch (2018/02/21) as possible. In the long term, as the beaming duration increases, initial excess or absence of intercepts tends to average out to the analytical value.\\

We conclude that our results yield no obvious dependence of intercept fraction on beaming duration, launch site, or target, as evidenced by the values in Tabs. \ref{tab:moon}, \ref{tab:mars}, and \ref{tab:pluto} (except the South Pole tends to have a higher intercept fraction). As a result, we can conclude that it is indeed feasible to track a target in the sky for extended periods of time with negligible intercept time with active satellites, given the assumptions detailed in the method section. This appears to hold true for all scenarios that were simulated. Although the beam may have to be gated off many times during the period of beaming in the long-term simulations (Mars, Moon, and Pluto), in general, the interception time is low, such that the beam can be turned on shortly after the intercepting satellite has passed. Furthermore, under the assumption that the intercept fraction is proportional to the number of satellites in orbit, the amount of satellites will need to increase by two to three orders of magnitude before significantly impacting the transmission time of these missions.

\subsection{Sidelobe Intercepts}

While the beam considered in this paper is assumed to be a single cylinder in profile, any real beam will have power outside this cylinder. This is often referred to as "sidelobe" structure. The detailed sidelobe structure will be important as the power levels are high enough that even relatively low level sidelobes need to be considered. For example a -30 db sidelobe with a 100 GW main beam is 100 MW which is still a major consideration for interception. The actual beam shape will depend on the specifics of the optical system and would need to be known in detail, though typical sidelobes are largely contained within the sub aperture beam size for a phased array. Since the satellites of interest are virtually all within the ``near field'' of the beam, this will limit the severity of the problem. Once the detailed beam structure is known, the same strategy used in this paper can be applied. \\

\section{Discussion} 
\label{Discussion}

Simulations of various DE mission scenarios of interest from multiple locations yielded results that are consistent with our analytical predictions. The intercept fraction as a function of beam time follows the law of large numbers and asymptotically approaches the worst-case analytical value of $4.9 \times 10^{-4}$ for a 10~km array (see Eq. (\ref{wc int fraction})). The median interception fraction is calculated to be $5.6 \times 10^{-4}$ for all long duration simulations. We expect that this value is higher than our worst-case estimate because we assumed the beam would point in the zenith direction at all times, and therefore take the minimum path length through the region of space occupied by satellites. \\

In general, missions from the South Pole yielded a longer beaming time above the horizon, but no beaming time above an altitude of 30$^{\circ}$.  Simulations at the South Pole also tended to have a higher intercept fraction, likely due to the targets being near its horizon (besides Alpha Centauri, which yielded a much lower intercept fraction). An orbital debris-debris collision avoidance system researched by NASA suggests laser sites closer to the poles for maximum targeting efficiency \citep{mason2011orbital}. Since we have the inverse goal, and as backed up by our simulations, it appears that beaming sites away from the poles would be better suited for our purposes.

\subsection{Limitations}
\label{Limitations}

There are a number of limitations that must be considered during this study. Firstly, the full catalog of satellites only includes unclassified satellites, and hence all classified satellites are completely untracked by the simulation. Consequently, there may be intercepts with our beam that we do not expect if operating purely on TLE data from the Space Track database. However, this is unlikely to be a significant issue, as there is likely to be a relatively low number of these satellites compared to unclassified satellites. In addition, it is probable that in practice, when anticipating a real mission, the Laser Clearing House (LCH) would help us determine times where we may intercept a classified satellite, and therefore we would be able to take appropriate evasive action. \\

Secondly, as mentioned previously, the SGP4 orbit propagator is a simplified general perturbation model and has a number of inaccuracies. More recently, ``specialized'' propagators have been designed that use higher-order numerical integration with more detailed models of all known forces acting on a LEO object. For example, a specialized propagator, called SDP4, models the gravitational effects of the sun and moon, and specific sectoral and tesseral Earth harmonics which are important for half-day and one-day period orbits \citep{spacetrackReport,atmos}. \\

As alluded to previously, it is important to recognize that SGP4 accuracy, as the prediction date gets further from the epoch, decreases dramatically. In a study conducted by Osweiler, the range of position errors encountered measuring normal and cross-track directions (normal to the plane defined by the current position and velocity vectors) using TLE and SGP4 was at most 0.25~km. The maximum velocity error was  greater at around 2~km/s after ten days. In order to be conservative we used the larger estimate of uncertainty of position (up to 3~km) given by \citet{satuncertainty} and \citet{sgp4}. These errors are greater when TLE's are used for satellites which experience significant amounts of drag (those with an orbital altitude of less than 800~km, i.e., LEO satellites \citep{osweiler2006covariance}). Furthermore, the BSTAR term in the TLEs is sometimes underestimated by multiple orders of magnitude when there is even low solar activity, since SGP4 assumes an atmospheric model that does not vary with solar activity. However, this inaccuracy is likely to be much more severe for debris bodies, and much less so for satellites \citep{mason2011orbital}. \\

In addition, there are now even more accurate methods for orbital propagation. The state of the art methods for this use numerical integration with high-fidelity force models (for example high-order gravitational models). Accuracy is primarily a function of the force models that are included and how they are taken into account, as well as the orbital regime of the satellite. In regimes such as GPS, it is possible to obtain accuracy on the order of centimeters at the epoch, since force terms such as atmospheric drag become negligible. \\

The simulations do not take the elevation of the beaming sites into account, but rather assume that they are all at sea level. This assumption creates an inaccuracy in the relative positions of the beam and satellites, particularly if the location of the laser array is at high altitude and if the pointing altitude angle is low. \\

Furthermore, as mentioned previously, the accuracy of TLE-derived ephemeris decays by 1--3~km a day and hence is only truly accurate up to two weeks about the epoch of the particular TLE (\cite{rhodes2}). This means that the long running simulations undertaken in this study (3 month, 1--3 year simulations) will not be accurate with regards to the ephemeris generation.\\

However, our research goal was not to obtain an exceptionally precise ephemeris for every satellite; the goal was to estimate general probabilities for interceptions, and to this end, the accuracies achieved by the SGP4 propagator and \texttt{PyEphem} more than suffice, since it is expected that the general distribution of satellites in space will not differ significantly, even in the long term. For future targeting systems and real mission scenarios, a similar program would require a means of dynamically updating the TLE to continuously correct for this accuracy decay while running (for example a targeting system for beaming power to the target \citep{finkleman2014dilemma}). \\ 

Lastly, our simulations do not take into account aircraft, which could also in theory intercept our beam. However, NASA has a Range Flight Safety program whose goal is to protect the safety and health of the public during operations. As proposed missions are under the NASA Innovative Advanced Concepts (NIAC) initiative, NASA Range Flight Safety would likely set up restricted airspace to avoid the area of the laser for the beaming duration (since aircraft can be rerouted unlike satellites \citep{nasaRangeSafety}). 

\section{Conclusions}
\label{Conclusion}

This paper has discussed simulations of pointing directed energy beams at targets in the sky, from a given location on the Earth, to thoroughly investigate the likelihood of Earth-orbiting spacecraft intercepting the beam and the duration of such intercepts.  \\

A theoretical analysis using a simple physical model was performed a priori, in order to predict the intercept fraction, and this was calculated to be $3.9 \times 10^{-4}$. \\

Through simulations of various scenarios, using a wide range of input parameters, such as beaming target, laser array site, beaming duration, and beam diameter, the results are consistent with this analysis. Indeed, it has been shown that there is a guarantee that we can find a window of time where there is exactly zero probability of intercepting an active Earth-orbiting satellite.\\

This research has widespread implications for the future of directed energy applications. We have shown it is feasible to point a directed energy beam at a celestial target for a wide range of durations, with negligible interception time with Earth-orbiting satellites. This opens up an array of possibilities for longer term DE mission proposals. In addition, accessible machinery now exists for future astronomers and others in the scientific community to predict satellite positions using TLE's.

\section*{Acknowledgements}
\label{Acknowledgements}

Funding for this program comes from NASA grants NIAC Phase I DEEP-IN – 2015 NNX15AL91G and NASA NIAC Phase II DEIS – 2016 NNX16AL32G and the NASA California Space Grant NASA NNX10AT93H as well as a generous gift from the Emmett and Gladys W. fund. PML acknowledges support as part of the advisory board and as part of the executive committee on Breakthrough Starshot.

\begin{appendix}    %>>>> this command starts appendixes

\section{NORAD Orbit propagator derivation}
\label{Appendix}

In this appendix, we outline the full derivation of the position and velocity vectors using the relevant TLE orbital parameters, as given by \cite{spacetrackReport}.

\subsection{Physical and Mathematical Constants}

The values of the physical and mathematical symbols used in the derivation that follows, are defined below. $n_o, e_o, i_o, M_o, \omega_o, \Omega_o, \dot{n}_o,$ and $\ddot{n}_o$ are all described by TLEs.
\begin{quote}
$R_\oplus$ = The radius of Earth = 6378.135~km\\[12pt]
$n_o$ = the SGP type ``mean'' mean motion at epoch\\[12pt]
$e_o$ = the ``mean'' eccentricity at epoch\\[12pt]
$i_o$ = the ``mean'' inclination at epoch\\[12pt]
$M_o$ = the ``mean'' mean anomaly at epoch \\[12pt]
$\omega_o$ = the ``mean'' argument of perigee at epoch \\[12pt]
$\Omega_o$ = the ``mean'' longitude of ascending node at epoch \\[12pt]
$\dot{n}_o$ = \parbox[t]{3in}{the time rate of change of ``mean'' mean motion at epoch} \\[12pt]
$\ddot{n}_o$ = \parbox[t]{3in}{the second time rate of change of ``mean'' mean motion at epoch} \\[12pt]
$B^*$ = the SGP4 type drag coefficient \\[12pt]
$k_\mathrm{e}$ = \parbox[t]{3in}{$\sqrt{GM}$ where $G$ is Newton's universal
gravitational constant and $M$ is the mass of the Earth} \\[12pt]
$a_\mathrm{E}$ = the equatorial radius of the Earth \\[12pt]
$J_2$ = the second gravitational zonal harmonic of the Earth \\[12pt]
$J_3$ = the third gravitational zonal harmonic of the Earth \\[12pt]
$J_4$ = the fourth gravitational zonal harmonic of the Earth \\[12pt]
$(t-t_o)$ = time since epoch \\[12pt]
$k_2 = \frac12J_2a_\mathrm{E}{}^2$ \\[12pt]
$k_4 = -\frac38J_4a_\mathrm{E}{}^4$ \\[12pt]
$A_{3,0} = -J_3a_\mathrm{E}{}^3$ \\[12pt]
$q_o$ = parameter for the SGP4/SGP8 density function \\[12pt]
$s$ = parameter for the SGP4/SGP8 density function \\[12pt]
$B$ = \parbox[t]{3in}{$\frac12C_D\frac Am$, the ballistic coefficient for SGP8
where $C_D$ is a dimensionless drag coefficient and $A$ is the average cross-%
sectional area of the satellite of mass $m$} \\[12pt]
\end{quote}

\subsection{SGP4 derivation}

The SGP4 model uses the NORAD mean element sets described by TLE's.  The original mean motion ($n''_o$) and semimajor axis ($a''_o$) are
calculated by the following equations:
\[a_1=\left(\frac{k_e}{n_o}\right)^{\frac23}\]
\[\delta_1=\frac32\frac{k_2}{a_1{}^2}\frac{(3\cos^2i_o-1)}
{(1-e_o{}^2)^{\frac32}}\]
\[a_o=a_1\left(1-\frac13\delta_1-\delta_1{}^2
-\frac{134}{81}\delta_1{}^3\right)\]
\[\delta_o=\frac32\frac{k_2}{a_o{}^2}\frac{(3\cos^2i_o-1)}
{(1-e_o{}^2)^{\frac32}}\]
\[n''_o=\frac{n_o}{1+\delta_o}\]
\[n''_o=\frac{a_o}{1-\delta_o}.\]
If the perigee of the orbit is between 98~km and 156~km, the value of $s$ used in SGP4 is modified to be
\[s^*=a''_o(1-e_o)-s+a_\mathrm{E}\]
If the perigee is below 98~km, the value of $s$ is modified to be
\[s^*=20/\mbox{$R_\oplus$}+a_\mathrm{E}.\]
If $s$ is modified, then the value $(q_o-s)^4$ must be replaced
by
\[(q_o-s^*)^4=\left[[(q_o-s)^4]^{\frac14}+s-s^*\right]^4.\]
Using the appropriate values of $s$ and
$(q_o-s)^4$, the following constants are calculated
\[\theta=\cos i_o\]
\[\xi=\frac{1}{a''_o-s}\]
\[\beta_o=(1-e_o{}^2)^{\frac12}\]
\[\eta=a''_o e_o\xi\]
\[C_2=\begin{array}[t]{l}(q_o-s)^4\xi^4n''_o(1-\eta^2)^{-\frac72}
\left[a''_o\left(1+\frac32\eta^2+4e_o\eta+e_o\eta^3\right)\right.\\[12pt]
\left.+\frac32\frac{k_2\xi}{(1-\eta^2)}
\left(-\frac12+\frac32\theta^2\right)(8+24\eta^2+3\eta^4)\right]
\end{array}\]
\[C_1=B^*C_2\]
\[C_3=\frac{(q_o-s)^4\xi^5A_{3,0}n''_o a_E\sin i_o}{k_2e_o}\]
\[C_4=\begin{array}[t]{l}2n''_o(q_o-s)^4\xi^4a''_o\beta_o{}^2(1-\eta^2)^{-\frac72}
\biggl(\left[2\eta(1+e_o\eta)
+\frac12e_o+\frac12\eta^3\right]\\[12pt] -\frac{2k_2\xi}{a''_o(1-\eta^2)}
\left[3(1-3\theta^2)
\left(1+\frac32\eta^2-2e_o\eta-\frac12e_o\eta^3\right)\right.\\[12pt]
\left.+ \frac34(1-\theta^2)(2\eta^2-e_o\eta-e_o\eta^3)\cos 2\omega_o\right]\biggr)
\end{array}\]
\[C_5=2(q_o-s)^4\xi^4a''_o\beta_o{}^2(1-\eta^2)^{-\frac72}
\left[1+\frac{11}4\eta(\eta+e_o)+e_o\eta^3\right]\]
\[D_2=4a''_o\xi C_1{}^2\]
\[D_3=\frac43a''_o\xi^2(17a''_o+s)C_1{}^3\]
\[D_4=\frac23a''_o\xi^3(221a''_o+31s)C_1{}^4.\]

The following equations account for long-term effects due to atmospheric drag and gravity
\[M_{\mathrm{DF}}=\begin{array}[t]{l}M_o+\left[1+\frac{3k_2(-1+3\theta^2)}{2a''_o{}^2\beta_o{}^3}
\right.\\[12pt] \left.+\frac{3k_2{}^2(13-78\theta^2+137\theta^4)}{16a''_o{}^4\beta_o{}^7}\right]
n''_o(t-t_o)\end{array}\]
\[\omega_{\mathrm{DF}}=\begin{array}[t]{l}\omega_o+\left[-\frac{3k_2(1-5\theta^2)}{2a''_o{}^2\beta_o{}^4}
+\frac{3k_2{}^2(7-114\theta^2+395\theta^4)}{16a''_o{}^4\beta_o{}^8}\right.\\[12pt]
\left.+\frac{5k_4(3-36\theta^2+49\theta^4)}{4a''_o{}^4\beta_o{}^8}\right]n''_o(t-t_o)
\end{array}\]
\[\Omega_{\mathrm{DF}}=\begin{array}[t]{l}\Omega_o+\left[-\frac{3k_2\theta}{a''_o{}^2\beta_o{}^4}
+\frac{3k_2{}^2(4\theta-19\theta^3)}{2a''_o{}^4\beta_o{}^8}
\right.\\[12pt] \left.+\frac{5k_4\theta(3-7\theta^2)}{2a''_o{}^4\beta_o{}^8}\right]n''_o(t-t_o)\end{array}\]
\[\delta\omega=B^*C_3(\cos\omega_o)(t-t_o)\]
\[\delta M=-\frac23(q_o-s)^4B^*\xi^4\frac{a_\mathrm{E}}{e_o\eta}
[(1+\eta\cos M_{\mathrm{DF}})^3-(1+\eta\cos M_o)^3]\]
\[M_\mathrm{p}=M_{\mathrm{DF}}+\delta\omega+\delta M\]
\[\omega=\omega_{\mathrm{DF}}-\delta\omega-\delta M\]
\[\Omega=\Omega_{\mathrm{DF}}-\frac{21}2\frac{n''_o k_2\theta}{a''_o{}^2\beta_o{}^2}
C_1(t-t_o)^2\]
\[e=e_o-B^*C_4(t-t_o)-B^*C_5(\sin M_\mathrm{p}-\sin M_\mathrm{o})\]
\[a=a''_o[1-C_1(t-t_o)-D_2(t-t_o)^2-D_3(t-t_o)^3-D_4(t-t_o)^4]^2\]
\[\L=\begin{array}[t]{l}
M_\mathrm{p}+\omega+\Omega+n''_o\left[\frac32C_1(t-t_o)^2+(D_2+2C_1{}^2)(t-t_o)^3\right.\\[12pt]
+\frac14(3D_3+12C_1D_2+10C_1{}^3)(t-t_o)^4\\[12pt]
\left.+\frac15(3D_4+12C_1D_3+6D_2{}^2+30C_1{}^2D_2+15C_1{}^4)(t-t_o)^5\right]
\end{array}\]
\[\beta=\sqrt{(1-e^2)}\]
\[n=k_\mathrm{e}\bigg/a^{\frac32}\]
where $(t-t_o)$ is the time that has elapsed from epoch. When the epoch perigee
height is under 220~km, the equations for $a$ and $\L$ are
truncated after the $C_1$ term, and terms involving $C_5$, $\delta\omega$,
and $\delta M$ are ignored.

The long-period periodic terms are added
\[a_{\mathrm{xN}}=e\cos\omega\]
\[\L_L=\frac{A_{3,0}\sin i_o}{8k_2a\beta^2}(e\cos\omega)
\left(\frac{3+5\theta}{1+\theta}\right)\]
\[a_{yNL}=\frac{A_{3,0}\sin i_o}{4k_2a\beta^2}\]
\[\L_\mathrm{T}=\L+\L_\mathrm{L}\]
\[a_{\mathrm{yN}}=e\sin\omega+a_{\mathrm{yNL}}.\]

Kepler's equation for $(E+\omega)$ is solved by defining
\[U=\L_\mathrm{T}-\Omega\]
and using the iteration equation
\[(E+\omega)_{i+1}=(E+\omega)_i+\Delta(E+\omega)_i\]
with
\[\Delta(E+\omega)_i =
\frac{U-a_{y\mathrm{N}}\cos(E+\omega)_i+a_{x\mathrm{N}}\sin(E+\omega)_i-(E+\omega)_i}
{-a_{y\mathrm{N}}\sin(E+\omega)_i-a_{x\mathrm{N}}\cos(E+\omega)_i+1}\]
and
\[(E+\omega)_1=U.\]

Preliminary quantities needed for short-period periodics are calculated using the following equations.
\[e\cos E=a_{x\mathrm{N}}\cos(E+\omega)+a_{x\mathrm{Y}}\sin(E+\omega)\]
\[e\sin E=a_{x\mathrm{N}}\sin(E+\omega)-a_{y\mathrm{N}}\cos(E+\omega)\]
\[e_L=(a_x{\mathrm{N}}{}^2+a_y{\mathrm{N}}{}^2)^{\frac12}\]
\[p_\mathrm{L}=a(1-e_\mathrm{L}{}^2)\]
\[r=a(1-e\cos E)\]
\[\dot{r}=k_e\frac{\sqrt{a}}re\sin E\]
\[r\dot{f}=k_\mathrm{e}\frac{\sqrt{p_L}}r\]
\[\cos u=\frac{a}r\left[\cos(E+\omega)-a_{x\mathrm{N}}+
\frac{a_{y\mathrm{N}}(e\sin E)}{1+\sqrt{1-e_\mathrm{L}{}^2}}\right]\]
\[\sin u=\frac{a}r\left[\sin(E+\omega)-a_{y\mathrm{N}}-
\frac{a_{x\mathrm{N}}(e\sin E)}{1+\sqrt{1-e_\mathrm{L}{}^2}}\right]\]
\[u=\tan^{-1}\left(\frac{\sin u}{\cos u}\right)\]
\[\Delta r=\frac{k_2}{2p_\mathrm{L}}(1-\theta^2)\cos 2u\]
\[\Delta u=-\frac{k_2}{4p_\mathrm{L}{}^2}(7\theta^2-1)\sin 2u\]
\[\Delta\Omega=\frac{3k_2\theta}{2p_L{}^2}\sin 2u\]
\[\Delta i=\frac{3k_2\theta}{2p_\mathrm{L}{}^2}\sin i_o\cos 2u\]
\[\Delta\dot{r}=-\frac{k_2n}{p_\mathrm{L}}(1-\theta^2)\sin 2u\]
\[\Delta r\dot{f}=\frac{k_2n}{p_\mathrm{L}}\left[(1-\theta^2)\cos 2u-\frac32(1-3
\theta^2)\right]\]

The short-period periodics are added to produce the following osculating quantities
\[r_k=r\left[1-\frac32k_2\frac{\sqrt{1-e_\mathrm{L}{}^2}}{p_\mathrm{L}{}^2}(3\theta^2-1)\right]
+\Delta r\]
\[u_k=u+\Delta u\]
\[\Omega_k=\Omega+\Delta\Omega\]
\[i_k=i_o+\Delta i\]
\[\dot{r}_k=\dot{r}+\Delta\dot{r}\]
\[r\dot{f}_k=r\dot{f}+\Delta r\dot{f}.\]
Unit orientation vectors are then calculated by
\[{\textit{\textbf{U}}}={\textit{\textbf{M}}}\sin u_k+{\textit{\textbf{N}}}\cos u_k\]
\[{\textit{\textbf{V}}}={\textit{\textbf{M}}}\cos u_k-{\textit{\textbf{N}}}\sin u_k\]
where
\[{\textit{\textbf{M}}}=\left\{\begin{array}{l}
                          M_x=-\sin\Omega_k\cos i_k\\
                          M_y=\cos\Omega_k\cos i_k\\
                          M_z=\sin i_k
                         \end{array}\right\}\]
\[{\textit{\textbf{N}}}=\left\{\begin{array}{l}
                          N_x=\cos\Omega_k\\
                          N_y=\sin\Omega_k\\
                          N_z=0
                         \end{array}\right\}.\]

Position is given by
\[{\textit{\textbf{r}}}=r_k{\textit{\textbf{U}}}\]
and velocity by
\[\dot{{\textit{\textbf{r}}}}=\dot{r}_k{\textit{\textbf{U}}}+(r\dot{f})_k{\textit{\textbf{V}}}.\]

\end{appendix}

%\section*{References}

\bibliography{report}

\begin{thebibliography}{25}
\expandafter\ifx\csname natexlab\endcsname\relax\def\natexlab#1{#1}\fi
\providecommand{\url}[1]{\texttt{#1}}
\providecommand{\href}[2]{#2}
\providecommand{\path}[1]{#1}
\providecommand{\DOIprefix}{doi:}
\providecommand{\ArXivprefix}{arXiv:}
\providecommand{\URLprefix}{URL: }
\providecommand{\Pubmedprefix}{pmid:}
\providecommand{\doi}[1]{\href{http://dx.doi.org/#1}{\path{#1}}}
\providecommand{\Pubmed}[1]{\href{pmid:#1}{\path{#1}}}
\providecommand{\bibinfo}[2]{#2}
\ifx\xfnm\relax \def\xfnm[#1]{\unskip,\space#1}\fi
%Type = Inproceedings
\bibitem[{Bible et~al.(2013)Bible, Johansson, Hughes and Lubin}]{pml2013}
\bibinfo{author}{Bible, J.}, \bibinfo{author}{Johansson, I.},
  \bibinfo{author}{Hughes, G.B.}, \bibinfo{author}{Lubin, P.M.},
  \bibinfo{year}{2013}.
\newblock \bibinfo{title}{Relativistic propulsion using directed energy}, in:
  \bibinfo{booktitle}{Nanophotonics and Macrophotonics for Space Environments
  VII}, \bibinfo{organization}{International Society for Optics and Photonics}.
  p. \bibinfo{pages}{887605}.
%Type = Inproceedings
\bibitem[{Brashears et~al.(2015)Brashears, Lubin, Hughes, McDonough, Arias,
  Lang, Motta, Meinhold, Batliner, Griswold et~al.}]{pml2015}
\bibinfo{author}{Brashears, T.}, \bibinfo{author}{Lubin, P.},
  \bibinfo{author}{Hughes, G.B.}, \bibinfo{author}{McDonough, K.},
  \bibinfo{author}{Arias, S.}, \bibinfo{author}{Lang, A.},
  \bibinfo{author}{Motta, C.}, \bibinfo{author}{Meinhold, P.},
  \bibinfo{author}{Batliner, P.}, \bibinfo{author}{Griswold, J.}, et~al.,
  \bibinfo{year}{2015}.
\newblock \bibinfo{title}{Directed energy interstellar propulsion of
  wafersats}, in: \bibinfo{booktitle}{Nanophotonics and Macrophotonics for
  Space Environments IX}, \bibinfo{organization}{International Society for
  Optics and Photonics}. p. \bibinfo{pages}{961609}.
%Type = Article
\bibitem[{Brouwer(1959)}]{brouwer1959solution}
\bibinfo{author}{Brouwer, D.}, \bibinfo{year}{1959}.
\newblock \bibinfo{title}{Solution of the problem of artificial satellite
  theory without drag}.
\newblock \bibinfo{journal}{The Astronomical Journal} \bibinfo{volume}{64},
  \bibinfo{pages}{378}.
%Type = Misc
\bibitem[{Deloach(2018)}]{nasaRangeSafety}
\bibinfo{author}{Deloach, R.}, \bibinfo{year}{2018}.
\newblock \bibinfo{title}{Nasa range flight safety program}.
\newblock \URLprefix \url{https://kscsma.ksc.nasa.gov/RangeSafety/overview}.
%Type = Article
\bibitem[{Finkleman(2014)}]{finkleman2014dilemma}
\bibinfo{author}{Finkleman, D.}, \bibinfo{year}{2014}.
\newblock \bibinfo{title}{The dilemma of space debris}.
\newblock \bibinfo{journal}{American Scientist} \bibinfo{volume}{102},
  \bibinfo{pages}{26}.
%Type = Article
\bibitem[{Hou et~al.(2016)Hou, Cai, Liu and Hou}]{Hou20161698}
\bibinfo{author}{Hou, L.}, \bibinfo{author}{Cai, Y.}, \bibinfo{author}{Liu,
  J.}, \bibinfo{author}{Hou, C.}, \bibinfo{year}{2016}.
\newblock \bibinfo{title}{Variable fidelity robust optimization of pulsed laser
  orbital debris removal under epistemic uncertainty}.
\newblock \bibinfo{journal}{Advances in Space Research} \bibinfo{volume}{57},
  \bibinfo{pages}{1698--1714}.
\newblock \URLprefix \url{https://doi.org/10.1016/j.asr.2015.12.003},
  \DOIprefix\doi{10.1016/j.asr.2015.12.003}. \bibinfo{note}{cited By 2}.
%Type = Inproceedings
\bibitem[{Hughes et~al.(2014)Hughes, Lubin, Griswold, Cook, Bozzini, O'Neill,
  Meinhold, Suen, Bible, Riley et~al.}]{pml2014}
\bibinfo{author}{Hughes, G.B.}, \bibinfo{author}{Lubin, P.},
  \bibinfo{author}{Griswold, J.}, \bibinfo{author}{Cook, B.},
  \bibinfo{author}{Bozzini, D.}, \bibinfo{author}{O'Neill, H.},
  \bibinfo{author}{Meinhold, P.}, \bibinfo{author}{Suen, J.},
  \bibinfo{author}{Bible, J.}, \bibinfo{author}{Riley, J.}, et~al.,
  \bibinfo{year}{2014}.
\newblock \bibinfo{title}{Optical modeling for a laser phased-array directed
  energy system}, in: \bibinfo{booktitle}{Nanophotonics and Macrophotonics for
  Space Environments VIII}, \bibinfo{organization}{International Society for
  Optics and Photonics}. p. \bibinfo{pages}{922603}.
%Type = Inproceedings
\bibitem[{Kelso(2007)}]{satuncertainty}
\bibinfo{author}{Kelso, T.}, \bibinfo{year}{2007}.
\newblock \bibinfo{title}{Validation of sgp4 and is-gps-200d against gps
  precision ephemerides}, in: \bibinfo{booktitle}{AAS/AIAA Space Flight
  Mechanics XVII}, \bibinfo{organization}{American Astronautical Society}.
\newblock \URLprefix \url{https://celestrak.com/publications/AAS/07-127/}.
%Type = Misc
\bibitem[{Kelso et~al.(1988)Kelso, Hoots and Roehrich}]{spacetrackReport}
\bibinfo{author}{Kelso, T.}, \bibinfo{author}{Hoots, F.},
  \bibinfo{author}{Roehrich, R.}, \bibinfo{year}{1988}.
\newblock \bibinfo{title}{Spacetrack report no. 3-models for propagation of
  norad element sets}.
%Type = Misc
\bibitem[{Kelso(2017)}]{kelso}
\bibinfo{author}{Kelso, T.S.}, \bibinfo{year}{2017}.
\newblock \bibinfo{title}{Celestrak}.
\newblock \URLprefix \url{https://celestrak.com/}.
%Type = Techreport
\bibitem[{Lane and Hoots(1979)}]{lane1979general}
\bibinfo{author}{Lane, M.H.}, \bibinfo{author}{Hoots, F.R.},
  \bibinfo{year}{1979}.
\newblock \bibinfo{title}{General perturbations theories derived from the 1965
  lane drag theory}.
\newblock \bibinfo{type}{Technical Report}. Aerospace Defense Command Peterson
  AFB CO Office of Astrodynamics.
%Type = Article
\bibitem[{Lejba et~al.(2018)Lejba, Suchodolski, Michałek, Bartoszak, Schillak
  and Zapaśnik}]{Lejba20182609}
\bibinfo{author}{Lejba, P.}, \bibinfo{author}{Suchodolski, T.},
  \bibinfo{author}{Michałek, P.}, \bibinfo{author}{Bartoszak, J.},
  \bibinfo{author}{Schillak, S.}, \bibinfo{author}{Zapaśnik, S.},
  \bibinfo{year}{2018}.
\newblock \bibinfo{title}{First laser measurements to space debris in poland}.
\newblock \bibinfo{journal}{Advances in Space Research} \bibinfo{volume}{61},
  \bibinfo{pages}{2609--2616}.
\newblock \URLprefix \url{https://doi.org/10.1016/j.asr.2018.02.033},
  \DOIprefix\doi{10.1016/j.asr.2018.02.033}. \bibinfo{note}{cited By 0}.
%Type = Article
\bibitem[{Lubin(2016)}]{pml2016}
\bibinfo{author}{Lubin, P.}, \bibinfo{year}{2016}.
\newblock \bibinfo{title}{A roadmap to interstellar flight}.
\newblock \bibinfo{journal}{Journal of the British Interplanetary Society –
  JBIS} \bibinfo{volume}{69}, \bibinfo{pages}{40--72}.
\newblock \URLprefix \url{http://arxiv.org/abs/1604.01356}.
%Type = Article
\bibitem[{Mason et~al.(2011)Mason, Stupl, Marshall and
  Levit}]{mason2011orbital}
\bibinfo{author}{Mason, J.}, \bibinfo{author}{Stupl, J.},
  \bibinfo{author}{Marshall, W.}, \bibinfo{author}{Levit, C.},
  \bibinfo{year}{2011}.
\newblock \bibinfo{title}{Orbital debris--debris collision avoidance}.
\newblock \bibinfo{journal}{Advances in Space Research} \bibinfo{volume}{48},
  \bibinfo{pages}{1643--1655}.
%Type = Misc
\bibitem[{McKeon(2016)}]{mckeon_2016}
\bibinfo{author}{McKeon, B.P.}, \bibinfo{year}{2016}.
\newblock \bibinfo{title}{Management of laser illumination of objects in
  space}.
\newblock \URLprefix \url{https://fas.org/irp/doddir/dod/i3100_11.pdf}.
%Type = Techreport
\bibitem[{Osweiler(2006)}]{osweiler2006covariance}
\bibinfo{author}{Osweiler, V.P.}, \bibinfo{year}{2006}.
\newblock \bibinfo{title}{Covariance estimation and autocorrelation of NORAD
  two-line element sets}.
\newblock \bibinfo{type}{Technical Report}. Air Force Inst Of Tech
  Wright-Patterson AFB OH School Of Engineering And Management.
%Type = Article
\bibitem[{Picone et~al.(2005)Picone, Emmert and Lean}]{atmos}
\bibinfo{author}{Picone, J.M.}, \bibinfo{author}{Emmert, J.T.},
  \bibinfo{author}{Lean, J.L.}, \bibinfo{year}{2005}.
\newblock \bibinfo{title}{Thermospheric densities derived from spacecraft
  orbits: Accurate processing of two-line element sets}.
\newblock \bibinfo{journal}{Journal of Geophysical Research: Space Physics}
  \bibinfo{volume}{110}.
\newblock \URLprefix \url{http://dx.doi.org/10.1029/2004JA010585},
  \DOIprefix\doi{10.1029/2004JA010585}. \bibinfo{note}{a03301}.
%Type = Misc
\bibitem[{Rhodes(2015)}]{sgp4}
\bibinfo{author}{Rhodes, B.}, \bibinfo{year}{2015}.
\newblock \bibinfo{title}{Standard general perturbation model 1.4}.
\newblock \URLprefix \url{https://pypi.python.org/pypi/sgp4/}.
%Type = Misc
\bibitem[{Rhodes(2008)}]{rhodes}
\bibinfo{author}{Rhodes, B.C.}, \bibinfo{year}{2008}.
\newblock \bibinfo{title}{Pyephem}.
\newblock \URLprefix \url{http://rhodesmill.org/pyephem/}.
%Type = Misc
\bibitem[{Rhodes(2018)}]{rhodes2}
\bibinfo{author}{Rhodes, B.C.}, \bibinfo{year}{2018}.
\newblock \bibinfo{title}{Pyephem}.
\newblock \URLprefix
  \url{https://rhodesmill.org/skyfield/earth-satellites.html}.
%Type = Article
\bibitem[{Rovetto and Kelso(2016)}]{rovetto2016preliminaries}
\bibinfo{author}{Rovetto, R.J.}, \bibinfo{author}{Kelso, T.},
  \bibinfo{year}{2016}.
\newblock \bibinfo{title}{Preliminaries of a space situational awareness
  ontology}.
\newblock \bibinfo{journal}{arXiv preprint arXiv:1606.01924} .
%Type = Article
\bibitem[{Southwood et~al.(1974)Southwood, May, Hassell and
  Conway}]{rstrategist}
\bibinfo{author}{Southwood, T.}, \bibinfo{author}{May, R.},
  \bibinfo{author}{Hassell, M.}, \bibinfo{author}{Conway, G.},
  \bibinfo{year}{1974}.
\newblock \bibinfo{title}{Ecological strategies and population parameters}.
\newblock \bibinfo{journal}{The American Naturalist} \bibinfo{volume}{108},
  \bibinfo{pages}{791--804}.
%Type = Article
\bibitem[{Swift et~al.(2013)Swift, Johnson, Morton, Crepp, Montet, Fabrycky and
  Muirhead}]{kepler}
\bibinfo{author}{Swift, J.J.}, \bibinfo{author}{Johnson, J.A.},
  \bibinfo{author}{Morton, T.D.}, \bibinfo{author}{Crepp, J.R.},
  \bibinfo{author}{Montet, B.T.}, \bibinfo{author}{Fabrycky, D.C.},
  \bibinfo{author}{Muirhead, P.S.}, \bibinfo{year}{2013}.
\newblock \bibinfo{title}{Characterizing the cool kois. iv. kepler-32 as a
  prototype for the formation of compact planetary systems throughout the
  galaxy}.
\newblock \bibinfo{journal}{The Astrophysical Journal} \bibinfo{volume}{764},
  \bibinfo{pages}{105}.
\newblock \URLprefix \url{http://stacks.iop.org/0004-637X/764/i=1/a=105}.
%Type = Inproceedings
\bibitem[{Vallado et~al.(2006)Vallado, Crawford, Hujsak and Kelso}]{revisiting}
\bibinfo{author}{Vallado, D.}, \bibinfo{author}{Crawford, P.},
  \bibinfo{author}{Hujsak, R.}, \bibinfo{author}{Kelso, T.},
  \bibinfo{year}{2006}.
\newblock \bibinfo{title}{Revisiting spacetrack report\# 3}, in:
  \bibinfo{booktitle}{AIAA/AAS Astrodynamics Specialist Conference and
  Exhibit}, p. \bibinfo{pages}{6753}.
%Type = Article
\bibitem[{Wang et~al.(2016)Wang, Zhang and Wang}]{Wang20161854}
\bibinfo{author}{Wang, C.}, \bibinfo{author}{Zhang, Y.}, \bibinfo{author}{Wang,
  K.}, \bibinfo{year}{2016}.
\newblock \bibinfo{title}{Impulse calculation and characteristic analysis of
  space debris by pulsed laser ablation}.
\newblock \bibinfo{journal}{Advances in Space Research} \bibinfo{volume}{58},
  \bibinfo{pages}{1854--1863}.
\newblock \URLprefix \url{https://doi.org/10.1016/j.asr.2016.07.018},
  \DOIprefix\doi{10.1016/j.asr.2016.07.018}. \bibinfo{note}{cited By 3}.

\end{thebibliography}

\end{document}